\pdfoutput=1

\documentclass[11pt,twoside,a4paper,cmspaper,final,collab]{cms-tdr}
\def\svnVersion{64972:64989}\def\cmsCernNo{2011-059}\def\cmsCernDate{\today}\def\cmsMessage{Submitted to the Journal of High Energy Physics}

\begin{document}\cmsNoteHeader{QCD-10-010}

\hyphenation{had-ron-i-za-tion}
\hyphenation{cal-or-i-me-ter}
\hyphenation{de-vices}
\RCS$Revision: 64989 $
\RCS$HeadURL: svn+ssh://alverson@svn.cern.ch/reps/tdr2/papers/QCD-10-010/trunk/QCD-10-010.tex $
\RCS$Id: QCD-10-010.tex 64989 2011-07-01 20:22:39Z alverson $

\def\lsim{\! \mathrel{\rlap{\lower4pt\hbox{\hskip1pt$\sim$}}
    \raise1pt\hbox{$<$}} \! }         
\def\gsim{\! \mathrel{\rlap{\lower4pt\hbox{\hskip1pt$\sim$}}
    \raise1pt\hbox{$>$}} \! }

\newcommand*{\py}{PYTHIA}

\newcommand*{\ptcut}{$p_T\!>\!0.5\,{\rm GeV/}c$}
\newcommand*{\etacut}{$|\eta|\!<\!2$}
\newcommand*{\delphi}{$\Delta\phi$}

\cmsNoteHeader{QCD-10-001.v3.1} 

\title{Measurement of the Underlying Event Activity \\
at the LHC  with $\sqrt{s} = 7~\TeV$ \\
and Comparison with  $\sqrt{s} = 0.9~\TeV$}       

\author[cern]{The CMS Collaboration}

\date{\today}

\abstract{
A measurement of the underlying activity in scattering processes with a
hard scale in the several \GeV\ region
is performed in proton-proton collisions at $\sqrt{s} = 0.9$ and $7 \TeV$, using data
collected by the CMS experiment at the LHC.
The production of charged particles with pseudorapidity \etacut\ and transverse momentum \ptcut\
is studied in the azimuthal region transverse to that
of the leading set of charged particles forming a track-jet.
A significant growth of the average multiplicity and scalar-$p_T$
sum of the particles in the transverse region is observed with increasing $p_T$
of the leading track-jet, followed by a much slower rise above a few \GeVc.
For track-jet $p_T$ larger than a few \GeVc, the activity in the transverse region
is approximately doubled with a centre-of-mass energy increase from $0.9$
to $7 \TeV$.
Predictions of several QCD-inspired models as implemented in {\sc pythia}
are compared to the data.
}

\hypersetup{%
pdfauthor={CMS Collaboration},%
pdftitle={Measurement of the Underlying Event Activity
at the LHC  with sqrt(s)= 7 TeV
and Comparison with  sqrt(s) = 0.9 TeV},%
pdfsubject={CMS},%
pdfkeywords={CMS, QCD, physics}}

\maketitle 

\section{Introduction}
\label{sec:Introduction}

In hadron-hadron scatterings, the ``underlying event'' (UE) is defined, in the presence
of a hard parton-parton scattering with large transverse momentum transfer, as any
hadronic activity that is additional to what can be attributed to the hadronization
of  partons involved in the hard scatter and to related initial and final state QCD radiation.
The UE activity is thus attributed to the hadronization of partonic constituents
that have undergone multiple parton interactions (MPI), as well as
to beam-beam remnants, concentrated along the beam direction.
Good understanding of UE properties is important for precision measurements
of standard model processes and the search for new physics at high energy.
Examples are the determination of the losses of events due to isolation criteria in lepton identification, or the computation of reconstruction efficiency for processes like H$ \to \gamma \gamma$ where the vertex is given by the underlying event.

The first measurement of UE activity at the LHC, with proton-proton centre-of-mass
energy $\sqrt{s} = 0.9\TeV$,  has been published by CMS~\cite{Khachatryan:2010pv}.
The present paper, which follows the same analysis procedure, reports on a measurement
at $\sqrt{s} = 7 \TeV$; new measurements at $0.9\TeV$ are also reported, with an event sample
$30$ times larger than that in Ref.~\cite{Khachatryan:2010pv}.
In this paper all measurements are fully corrected for detector effects. Details are given below.
The ATLAS collaboration has reported on measurements at
$\sqrt{s} = 0.9$ and $7\TeV$~\cite{Collaboration:2010fh}, using slightly different analysis procedures.

The UE activity at a given centre-of-mass energy is expected to increase with the hard scale in the interaction. Events with a harder scale are indeed expected to correspond, on average, to interactions with a smaller impact parameter, a feature which in turn should enhance MPI~\cite{Sjostrand:1986ep, Frankfurt:2010ea}. This increased activity is observed to reach a plateau for  high scales, corresponding to an MPI ``saturation" effect for impact parameters selected by a sufficiently hard leading interaction.
Conversely, for events with the same hard scale but taken at different values of the centre-of-mass energy, MPI activity is expected to increase with $\sqrt{s}$~\cite{Sjostrand:1986ep, Frankfurt:2010ea}. The present analysis is focused on measurements that can contribute to the understanding of the UE dynamics, through the comparison of events at the same $\sqrt{s}$ but with different hard scales, and the comparison of data with the same hard scale but with different values of $\sqrt{s}$.

To study the UE, it is convenient to refer to the difference in azimuthal angle,
$\Delta\phi$,
between the projections onto the plane perpendicular to the beam of the
directions of the hard scatter and of any hadron in the event.
With this method, the UE activity is made manifest  in the ``transverse'' region
with $60^\circ \!<\! |\Delta\phi|  \!<\! 120^\circ$, even though it cannot
in principle be uniquely separated from initial and final state radiation.
In this paper, the direction of the hard scatter is identified with that of the
leading ``track-jet'', i.e. the object with largest transverse momentum, $p_T$, formed using a jet algorithm applied to reconstructed tracks of particles above some minimum $p_T$ value in the event.
The leading track-jet $p_T$ is taken as defining the hard scale in the event.
An advantage of using a track-jet as a reference is that it is an
experimentally well-defined object, essentially free from pileup effects.
No attempt is made to refer to the corresponding parton-level objects, as this would result in additional model uncertainties. However, the track-jet is much closer  to the parton-level object than the leading track.
Finally, in the few \GeVc\ region, the value of the track-jet $p_T$ is better defined and more stable than for  calorimeter based jets, which suffer from large fluctuations.

The UE dynamics are studied through the comparison with data of models
implemented in Monte Carlo (MC) simulations adopting MPI.
The predictions of the models without MPI fail to reproduce the evolution of the UE observables with the scale of the interaction and with the centre-of-mass energy~\cite{Alekhin:2005dx}.
The predictions for inelastic events are provided here by several tunes of the
{\sc pythia} program, versions {\sc 6.420}~\cite{Sjostrand:1986ep,Sjostrand:2006za}
and {\sc 8.145}~\cite{Sjostrand:2007gs,Corke:2009pm}.
The pre-LHC D6T tune~\cite{Field:2008zz,rdf1} of {\sc pythia6}, which describes the
lower energy UA5 and Tevatron data, is a widely used reference that will
also be used for the present analysis.
The tunes DW~\cite{rdf1} and CW~\cite{Khachatryan:2010pv}, which were found to
describe best the data at $0.9 \TeV$~\cite{Khachatryan:2010pv}, will be discussed for the present $7 \TeV$ data.
The new {\sc pythia6} tune, Z1~\cite{Field:2010bc}, includes $p_T$ ordering of parton showers
and the new {\sc pythia} MPI model~\cite{Skands:2007zg}. It implements the results of the Professor tunes~\cite{Buckley:2009bj} considering the fragmentation and the coulor reconnection parameters of the AMBT1 tune~\cite{:2010ir}; preliminary CMS UE results at $7 \TeV$
have been used to tune the parameters governing the value and the $\sqrt{s}$ dependence of the transverse momentum cutoff that in {\sc pythia} regularizes the divergence of  the leading order scattering amplitude
as the final state parton transverse momentum $\hat{p}_T$ approaches $0$.
The tune Z2 is similar to Z1, except for the transverse momentum cutoff  at the nominal energy of  $\sqrt{s_0} = 1.8 \TeV$ which is decreased by $0.1 \GeVc$.
{\sc pythia8} also uses the new {\sc pythia} MPI model, which is interleaved with parton showering.
The new {\sc pythia8} version 4C~\cite{Corke:2010yf}  has also been tuned to the early LHC data.
The {\sc pythia8} model includes soft and hard diffraction~\cite{diffr-in-pythia_navin}, whereas only soft diffraction is included in {\sc pythia6}; the precise description of diffraction is, however, of little relevance for the present analyses since it has been checked that the diffractive contributions are strongly suppressed by the
trigger and event selection requirements, especially for large $p_T$ values of the leading track-jet.
The parton distribution functions (PDF) used to describe the protons are
the CTEQ6L1 set~\cite{Pumplin:2002vw} for D6T, Z2 and 4C, and CTEQ5L~\cite{Lai:1999wy} for the other simulations.

The outline of the paper is as follows.
Section~\ref{sec:Experimental} presents experimental details: brief detector description, data samples,
event and track selection, track-jet reconstruction, unfolding procedure and
systematic uncertainties.
Section~\ref{sec:Results} presents results on the transverse region dynamics:
hard-scale dependence and particle spectra at $\sqrt{s} = 7 \TeV$,
and centre-of-mass energy dependence of the transverse region dynamics.
Section~\ref{sec:Conclusions} summarizes the main results of the study and
draws conclusions.

\section{ Experimental Details}
\label{sec:Experimental}

A description of the CMS detector can be found in Ref.~\cite{JINST}.
The coordinate system has the origin at the nominal interaction point.
The $z$ axis is parallel to the anticlockwise beam direction; it defines the polar angle
$\theta$ and the pseudorapidity $\eta = -\ln (\tan(\theta / 2)  )$.
The azimuthal angle $\phi$ is measured in the plane transverse to the beam,
from the direction pointing to the centre of the LHC ring toward the upward
direction.
The pixel and silicon strip tracker, immersed in the uniform $3.8\unit{T}$ magnetic
field provided by a $6  \unit {m}$ diameter superconducting solenoid,
measures charged particle trajectories in the pseudorapidity range
$|\eta| \!<\! 2.5$.
The $p_T$ resolution for $1 \GeVc$ charged particles is between
$0.7\%$ at $\eta = 0$ and $2\%$ at $|\eta| = 2.5$.

For this analysis, the same selection conditions apply to events and
tracks at $0.9$ and $7\TeV$; these conditions are very similar to those
at $0.9\TeV$ in Ref.~\cite{Khachatryan:2010pv}.
Minimum bias events are triggered by requiring activity in both
Beam Scintillator Counters (BSC)~\cite{JINST,Bell:1204301}, in coincidence with signals  from
both beams in the Beam Pick-up Timing for eXperiments  (BPTX) devices~\cite{JINST,Aumeyr:2008mt}; low-$p_T$ track-jets are recorded with a prescaled minimum bias trigger.  At $7\TeV$, in order to enhance the acquisition of events with a harder scale and reduce statistical fluctuations, the analysis also uses single-jet triggers.

Selected events are required to contain one and only one primary vertex,
reconstructed in fits with more than four degrees of freedom,
with a $z$ coordinate within $10  \unit {cm}$ of the centre of the $4
\unit{cm}$-wide beam collision region.
Rejecting events with more than one primary vertex does not bias the final results, as was checked by comparing data with different pileup conditions, taken at low and high instantaneous luminosities.

Selected events are also required to contain a track-jet with  $p_{T}\!>\!1 \GeVc$,
reconstructed with pseudorapidity \etacut.
Track-jets are defined using the SISCone algorithm~\cite{Salam:2007xv}
as implemented in the FastJet package~\cite{Cacciari:2005hq} with a clustering
radius $R = \sqrt {(\Delta \phi)^2 + (\Delta \eta)^2 } = 0.5$.
Charged particles reconstructed in the tracker with \ptcut\ and
$|\eta|\!<\!2.5$ are used to define the track-jet; this $\eta$~range is wider than that
used for the UE analysis (\etacut ) in order to avoid a kinematic bias.

A track is selected for the UE analysis if it is consistent with the primary vertex and is
reconstructed in the pixel and silicon strip tracker with transverse momentum \ptcut\
and pseudorapidity \etacut.
A high-purity reconstruction algorithm is used, which keeps low levels of misreconstructed and
poorly reconstructed tracks~\cite{Khachatryan:2010pw}.
To decrease contamination by secondary tracks from decays of long-lived
particles and photon conversions,
the distance of closest approach between the track and the primary vertex is required
to be less than three times its (significantly non-Gaussian) estimated uncertainty,
both in the transverse plane and along the $z$-axis;
the uncertainty in the vertex position is also taken into account.
Poorly measured tracks are removed by requiring  $\sigma(p_T)/p_T \!<\! 5\%$,
where $\sigma(p_T)$ is the uncertainty on the $p_T$  measurement.
In the selected track sample with \etacut, these selections result in a background
level of  3\% : 1\% from $\rm{K_{S}^0}$ and  $\Lambda^{0}$ decay products
and 2\% from combinatorial background, as estimated using MC simulations.
The number of selected events at both centre-of-mass energies, for
track-jet $p_T\!>\!1$, $3$, and $20 \GeVc$, and the corresponding number
of selected tracks are given in Table~\ref{tab:selectEvents}.

\begin{table}
\caption
{
\label{tab:selectEvents}
Number of selected events and corresponding number of selected tracks, for three values of minimum
track-jet $p_T$ and for both $\sqrt{s} = 0.9$ and $7\TeV$.}
\begin{center}
\begin{tabular}{c|c|c|c|c|c|c}
      &   \multicolumn{3}{|c| } {$\sqrt{s} = 7\TeV$}  & \multicolumn{3}{|c } {$\sqrt{s} = 0.9\TeV$}  \\
\hline
leading track-jet $p_{T}\!>$
  & $1 \GeVc$  & $3 \GeVc$  & $20 \GeVc$ & $1 \GeVc$  & $3 \GeVc$  & $20 \GeVc$ \\
\hline
No. selected events ($ \times 10^3$)  &  18 543    &   6 674   &  19    &  5 140     & 783     & 0.25 \\
No. selected tracks ($ \times 10^3$)  &  202 952  &  120 945 &  638  &  33 743  & 9 001  & 5.8 \\
\end{tabular}
\end{center}
\end{table}

The distributions presented below are fully corrected for detector effects.
An iterative unfolding technique~\cite{D'Agostini:1994zf} is used, except for some
cases that will be detailed below.
The {\sc pythia6} MC with Z2 tune was used to correct the experimental distributions,
while Z1, D6T and the default configuration of {\sc pythia 8.135} (``tune~1'') were
used for cross-checks and systematic uncertainty estimates.
The detector response was simulated in detail using  the {\sc GEANT4}
package~\cite{Agostinelli:2002hh}, and simulated events were processed and reconstructed
in the same manner as collision data.
The simulations were found to give a very good description of all features related to
detector performance that are relevant to this analysis.
The unfolding procedure was tested using MC events, by comparing the genuine distributions for generated hadrons with the distributions obtained, after unfolding, from reconstructed tracks.

Systematic uncertainties on the corrected data have been studied in detail.
They correspond essentially to the uncertainties described in~\cite{Khachatryan:2010pv} taking into account the progress reported in Ref.~\cite{Khachatryan:2010pw}.
They include the implementation in the simulation of vertex and track selection criteria,
tracker alignment and tracker material content, background contamination from
 $\rm{K_{S}^0}$ and $\Lambda^{0}$ production, trigger conditions, run-to-run variations of tracker and beam conditions, including the effect of pileup, and the effect of limited samples.

Using as MC input the Z1 simulation, which gives the best description of data,
unfolding procedures were performed using both the Z2 and tune~1 models; the maximum discrepancies with the MC input were taken as systematic uncertainties.

Systematic uncertainties are largely independent of one another, but they are
correlated among data points in each experimental distribution.
They are added in quadrature to statistical uncertainties and represented in all figures.

\section{Underlying Event in the Transverse Region }
\label{sec:Results}

The hadronic activity at $7 \TeV$ in the transverse region, for charged particles with \ptcut,
\etacut, and $60^\circ\!<\! |\Delta\phi| \!<\! 120^\circ$,
is first presented as a function of the leading track-jet $p_T$ (Section~\ref{sec:profiles}).
Multiplicity and transverse momentum distributions  are then reported for two
minimal values, $3$ and $20 \GeVc$, of the leading track-jet $p_T$  (Section~\ref{sec:spectra}).
Results at the two centre-of-mass energies, $\sqrt{s} = 0.9 $
and $ 7 \TeV$, are finally compared  (Section~\ref{sec:sqrt-s-dependence}).
Predictions from the various {\sc pythia} models are compared to the corrected data.

\subsection{Hard-scale dependence}
\label{sec:profiles}

\begin{figure}[htbp]
\begin{center}
\includegraphics[angle=90, width=0.32 \textwidth]{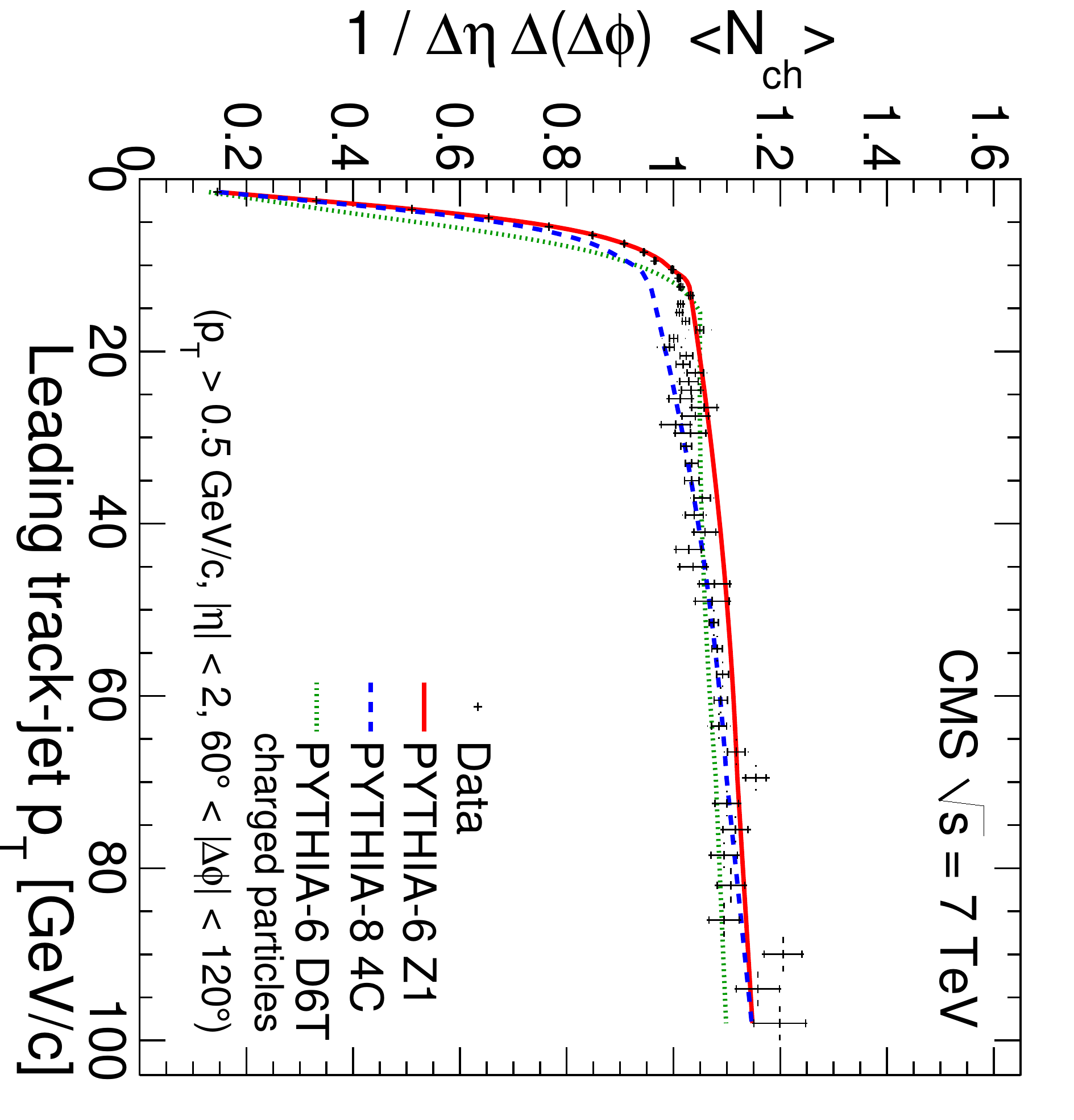}
\includegraphics[angle=90, width=0.32 \textwidth]{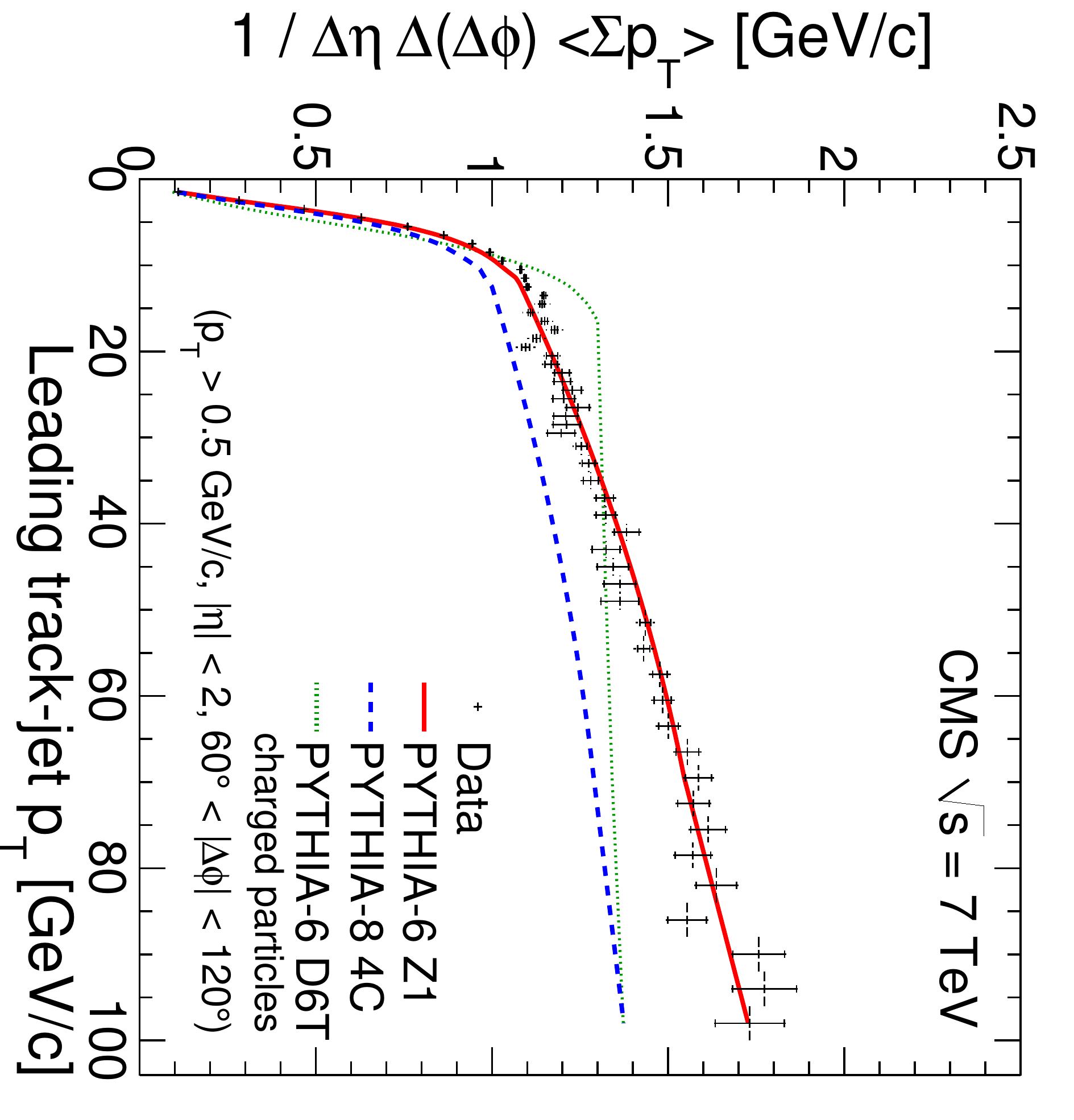}
\includegraphics[angle=90, width=0.32 \textwidth]{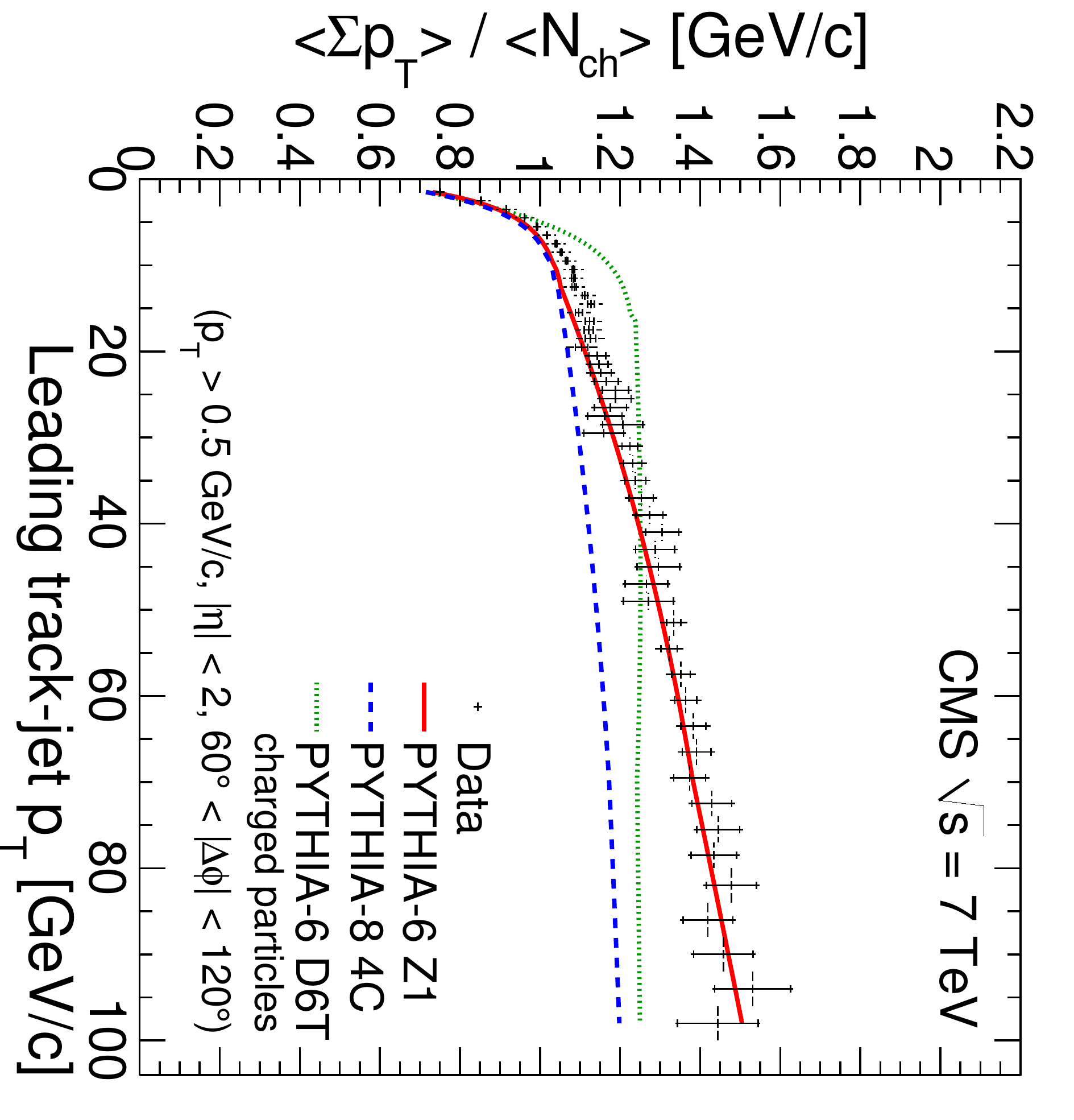}
\end{center}
\caption{
Fully corrected measurements of charged particles with \ptcut\ and \etacut\
in the transverse region, $60^\circ\!<\! |\Delta\phi| \!<\! 120^\circ$,
as a function of the $p_T$  of the leading track-jet:
(left) average multiplicity per unit of pseudorapidity  and per radian;
(centre) average scalar $\sum p_T$ per unit of pseudorapidity  and per radian;
(right) ratio of the average scalar $\sum p_T$ and the average multiplicity.
Predictions of three {\sc pythia} tunes are compared to the data.
\label{fig:Fig-3}}
\end{figure}

Figure~\ref{fig:Fig-3} presents the average multiplicity
and the average scalar momentum sum
in the transverse region, as a function of the leading track-jet $p_T$.
For these distributions, full unfolding was performed in both the track-jet $p_T$ and the
studied variable, for leading track-jet $p_T$ up to $20 \GeVc$.
Bin-by-bin corrections were used at higher values of the leading track-jet $p_T$ where the $p_T$ dependence of the studied variables is small.

The horizontal error bars indicate the bin size; the vertical inner error bars indicate the statistical uncertainties affecting the measurements; the outer error bars represent the statistical and systematic uncertainties added in quadrature; statistical uncertainties dominate at large values of the hard scale.
The same conventions and considerations apply throughout this paper.

Two components are visible for both observables in Fig.~\ref{fig:Fig-3}:
a fast rise for $p_T \ \lsim \ 8 \GeVc$, attributed mainly to the increase of
MPI activity, followed by a plateau-like region with nearly
constant average number of selected particles and a slow increase of
$\Sigma p_{T}$.
A similar structure is observed at $0.9 \TeV$
(see Ref.~\cite{Khachatryan:2010pv} and Fig.~\ref{fig:three-prime} below),
the fast rise being limited in
that case to the region with leading track-jet $p_T \ \lsim \ 4 \GeVc$.
All {\sc pythia} models predict such a distinct change of the amount of activity in the
transverse region as a function of the leading track-jet $p_T$.

These evolutions result in a slow but continuous increase of the average $p_T$ of the selected
particles for leading track-jet $p_T$ above a few $\GeVc$.
This is observed in Fig.~\ref{fig:Fig-3} (right plot), which is obtained from the
ratio of the two profile distributions, with the relative uncertainties conservatively summed in quadrature. (A similar behaviour of the ratio can be deduced at $0.9 \TeV$).

\begin{figure}[htbp]
\begin{center}
\includegraphics[angle=90, width=0.32 \textwidth]{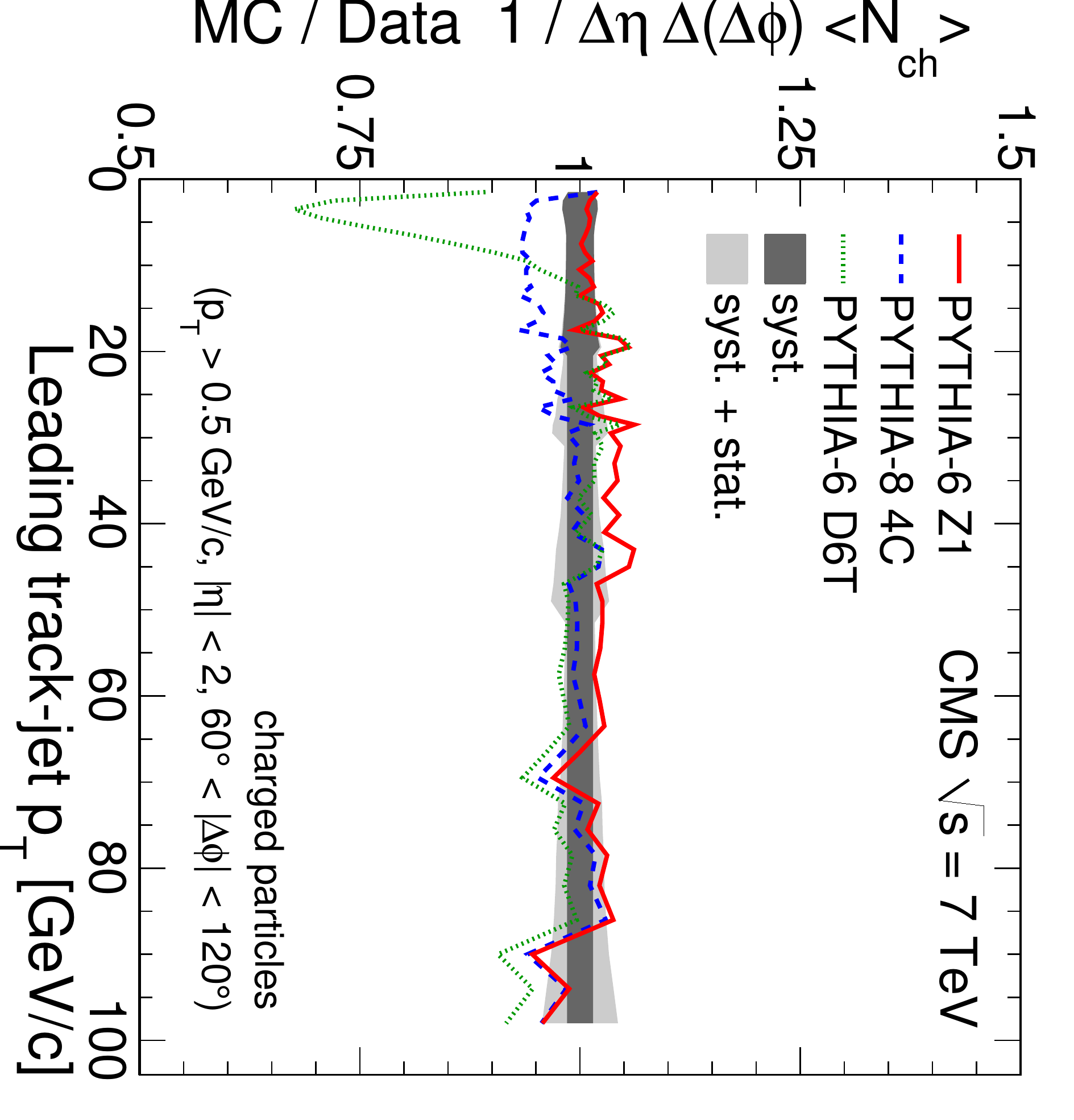}
\includegraphics[angle=90, width=0.32 \textwidth]{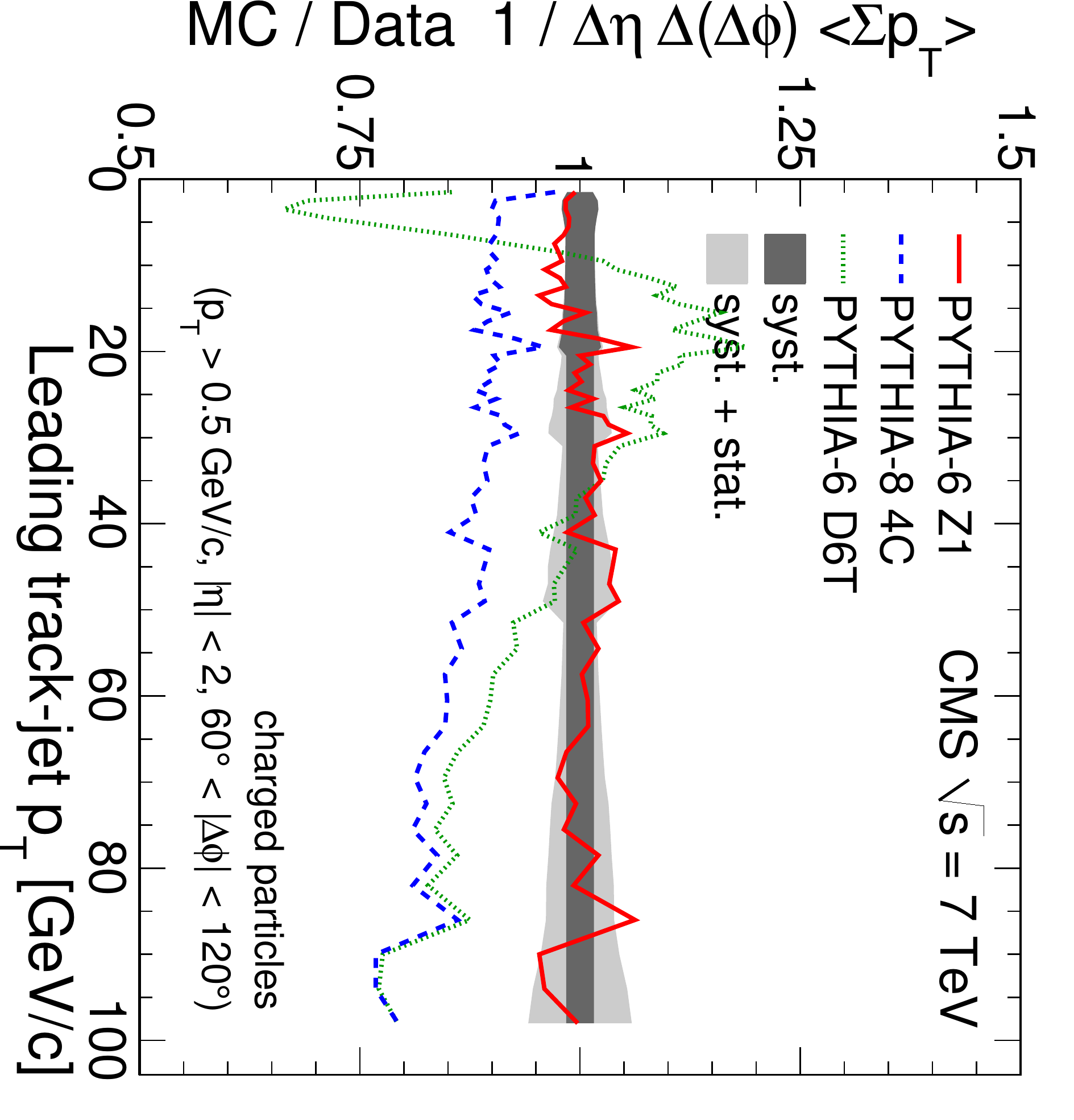}
\includegraphics[angle=90, width=0.32 \textwidth]{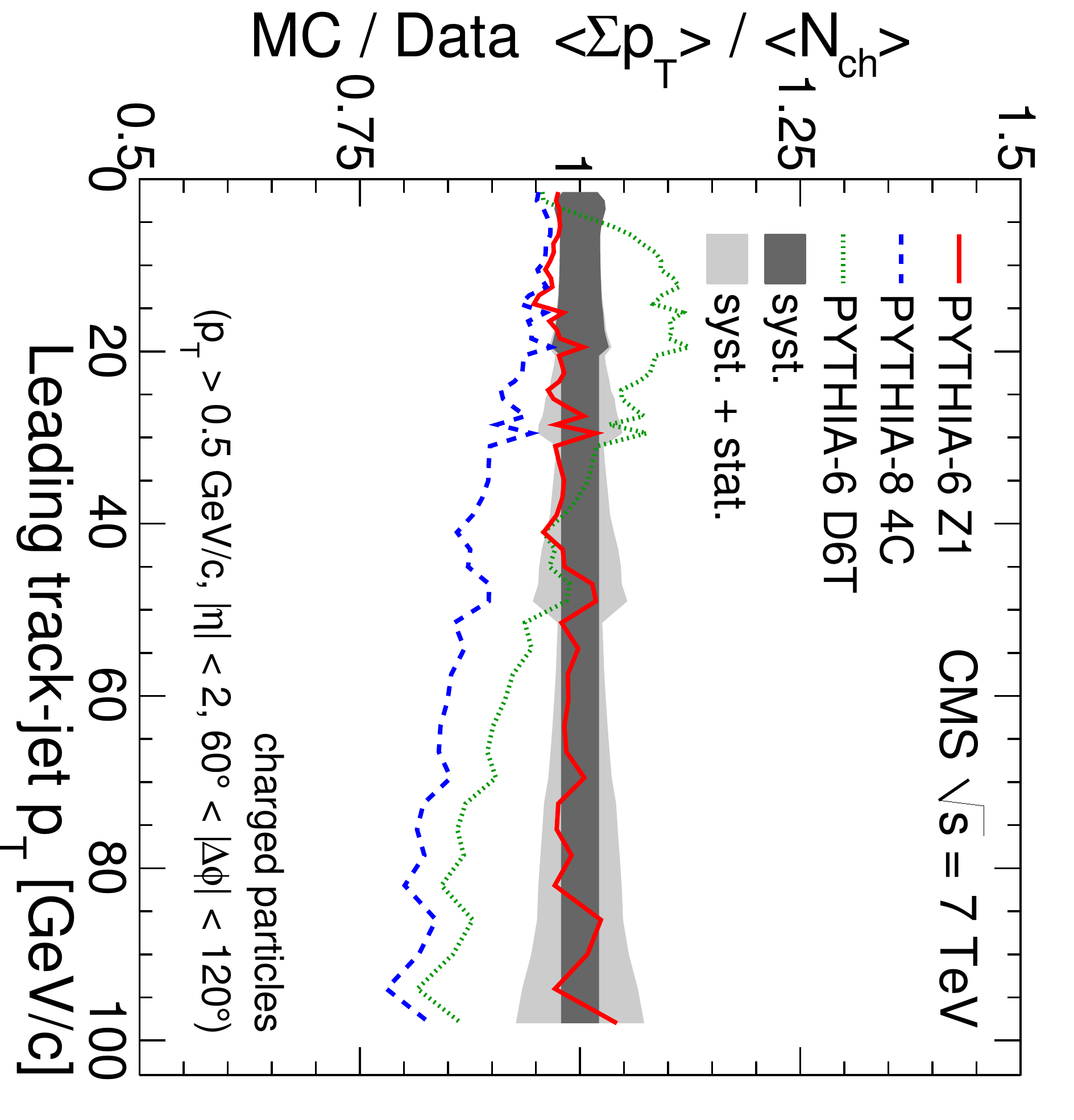}
\end{center}
\caption{
Ratios, as a function of the leading track-jet $p_T$,
of three MC predictions to the fully corrected measurements
of charged particles with \ptcut\ and \etacut\ in the transverse region,
$60^\circ\!<\! |\Delta\phi| \!<\! 120^\circ$
(cf.~Fig.~\ref{fig:Fig-3}):
(left) average multiplicity;
(centre) average scalar $\sum p_T$;
(right) ratio of the average scalar $\sum p_T$ and the average multiplicity.
The inner bands correspond to the systematic uncertainties and the outer bands
to the total experimental uncertainties (statistical and systematic uncertainties
added in quadrature).
\label{fig:ratiosToMany-1} }
\end{figure}

The information on the quality of the data description by the different models
is summarized in Fig.~\ref{fig:ratiosToMany-1},
which presents the ratio of the MC predictions to the measurements for the
observables shown in Fig.~\ref{fig:Fig-3}.
Statistical fluctuations in the data induce correlated fluctuations for the various
MC/data ratios.
Variations in the error bands are related to the unfolding procedures, and to the
different sets of data collected with different triggers.

The description provided by Z1 is very good for both  the average multiplicity
and the average scalar momentum sum, over the full
leading track-jet $p_T$ range. For the {\sc pythia8} 4C tune, in the region with track-jet $p_T$  $ < 20 \GeVc$ the predictions are below the data by  5\% (10\%) for the average charged multiplicity (average scalar $\sum p_T$); for larger track-jet $p_T$ values, the average charged multiplicity is well described but  the average $\sum p_T$ is increasingly underestimated, by up to 20\%. This confirms observations reported in Ref.~\cite{Buckley:2011ms}.
The predictions of the Z2 tune (not shown in this paper) reveal similar trends as for {\sc pythia8} 4C with, however, a more uniform and more limited underestimate ($\ \lsim \ 10\%$) of the average $\sum p_T$.

As illustrated by D6T, the predictions of older {\sc pythia6} tunes are significantly below the data in the region characterized by the fast rise of the observables (track-jet $p_T$ $\lsim \ 8 \GeVc$); in the ``saturation'' region, tune D6T provides a good description of the average multiplicity but the average $\sum p_T$ is largely underestimated for track-jet $p_T$ $ > 40 \GeVc$. In this region, DW predictions are lower than for D6T, and CW even lower, which reflects the different values of the cutoff transverse momentum and its $\sqrt{s}$
dependence~\cite{Khachatryan:2010pv} (CW and the DW predictions are not shown in this paper).

The comparison with data of the MC predictions for the ratio of the average $\sum p_T$
and the average multiplicity is shown
in Fig.~\ref{fig:ratiosToMany-1} (right).
The absolute value is described within 5\% and the hard-scale dependence is well described by Z1. The {\sc pythia8} 4C and Z2 predictions agree with Z1 in the rising region, but the normalization is up to 10\% and 20\% lower in the ``saturation'' region, respectively.
For D6T, the ratio is overestimated below $30 \GeVc$, and underestimated above $30 \GeVc$.

\subsection{Multiplicity and transverse momentum distributions}
\label{sec:spectra}

\begin{figure}[htbp]
\begin{center}
\includegraphics[angle=90, width=0.32 \textwidth]{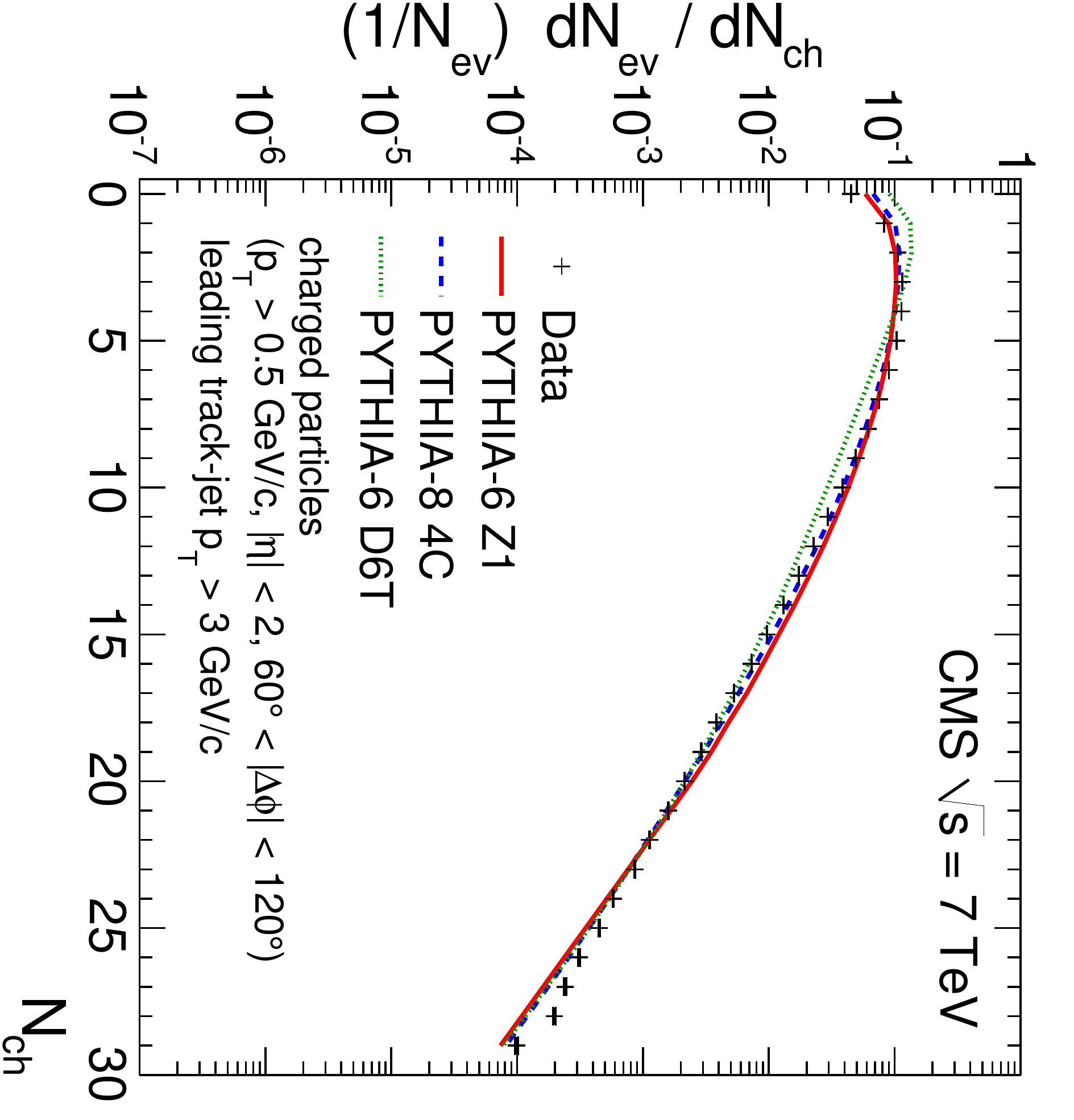}
\includegraphics[angle=90, width=0.32 \textwidth]{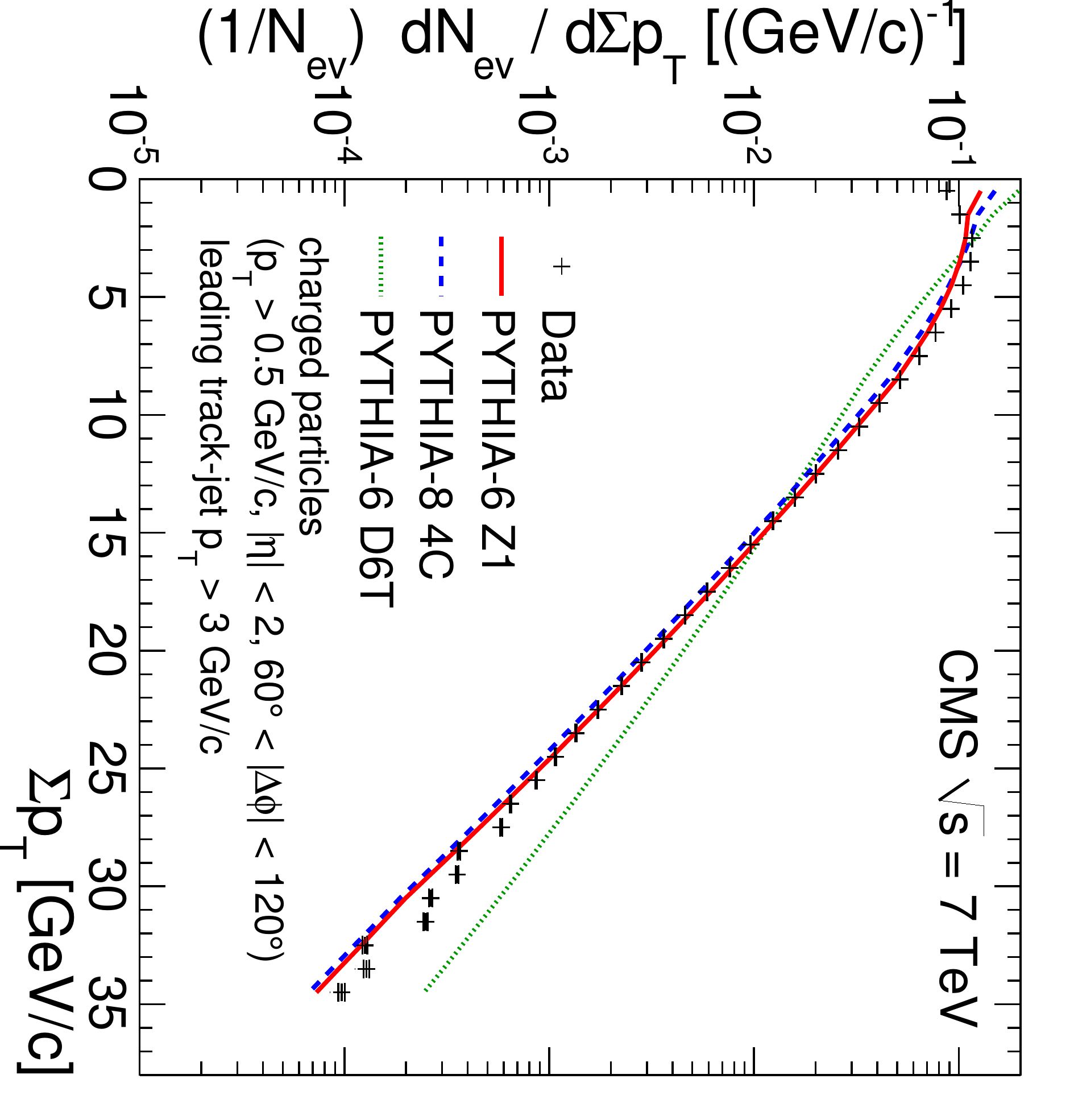}
\includegraphics[angle=90, width=0.32 \textwidth]{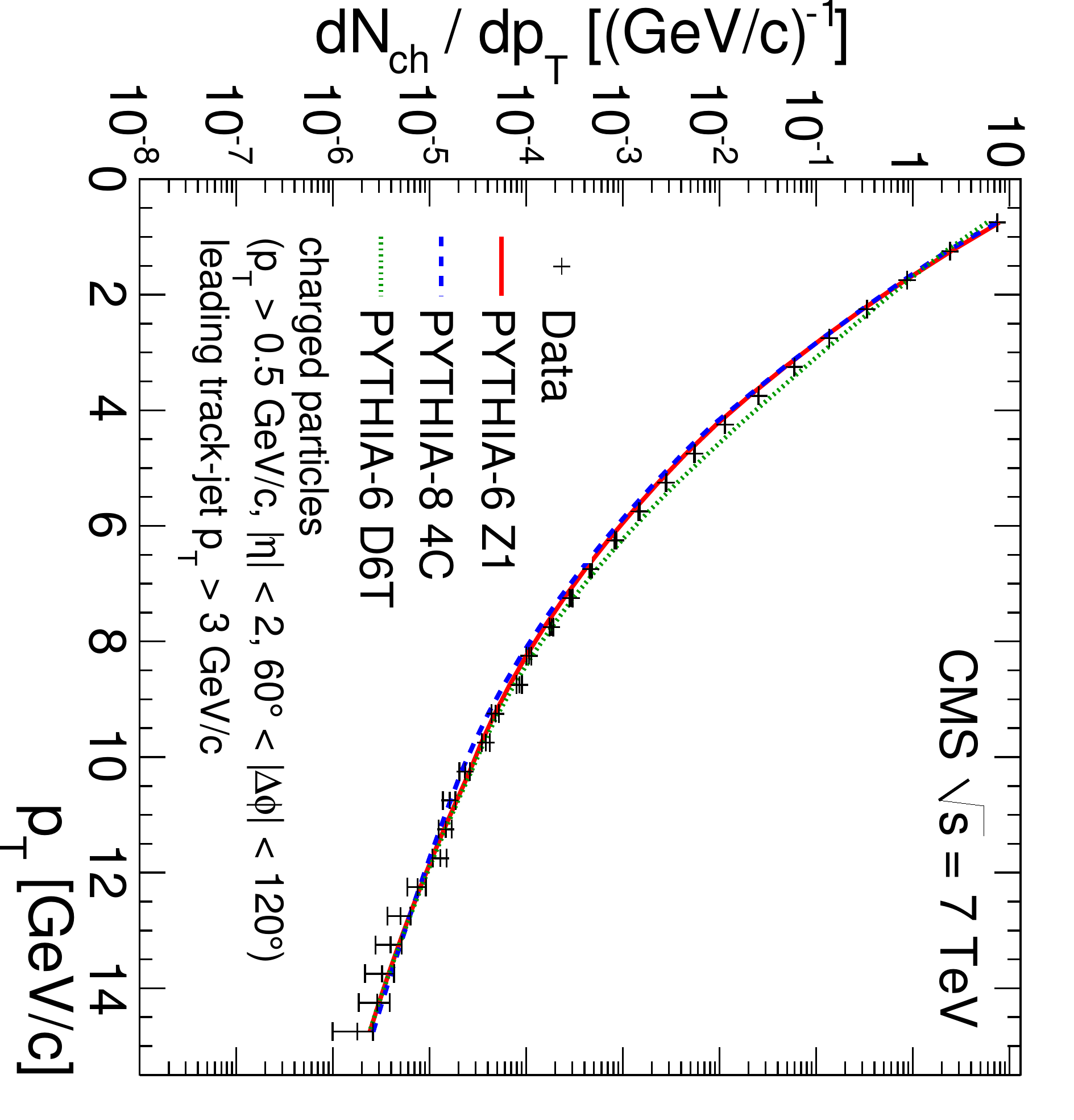} \\
\includegraphics[angle=90, width=0.32 \textwidth]{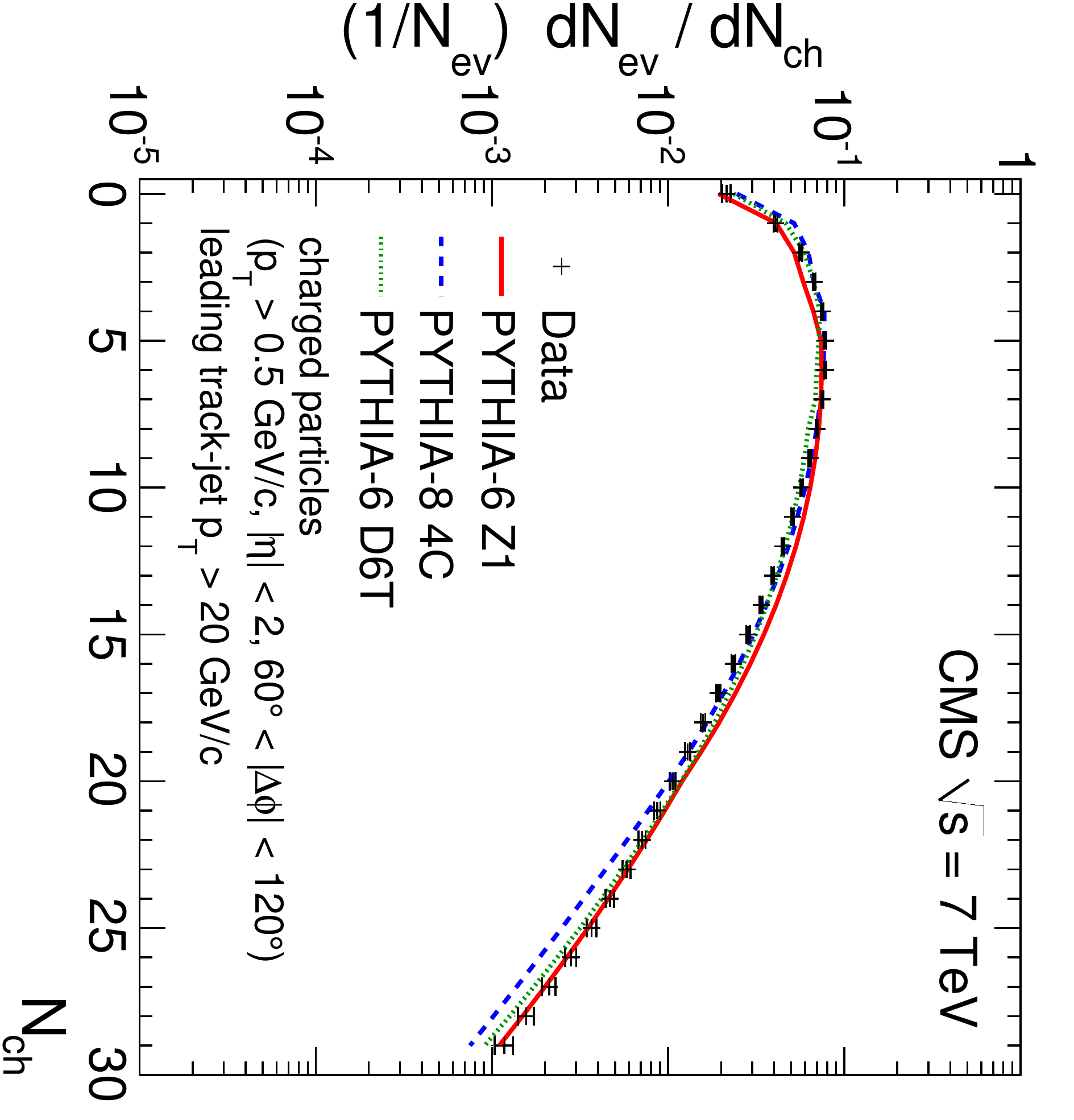}
\includegraphics[angle=90, width=0.32 \textwidth]{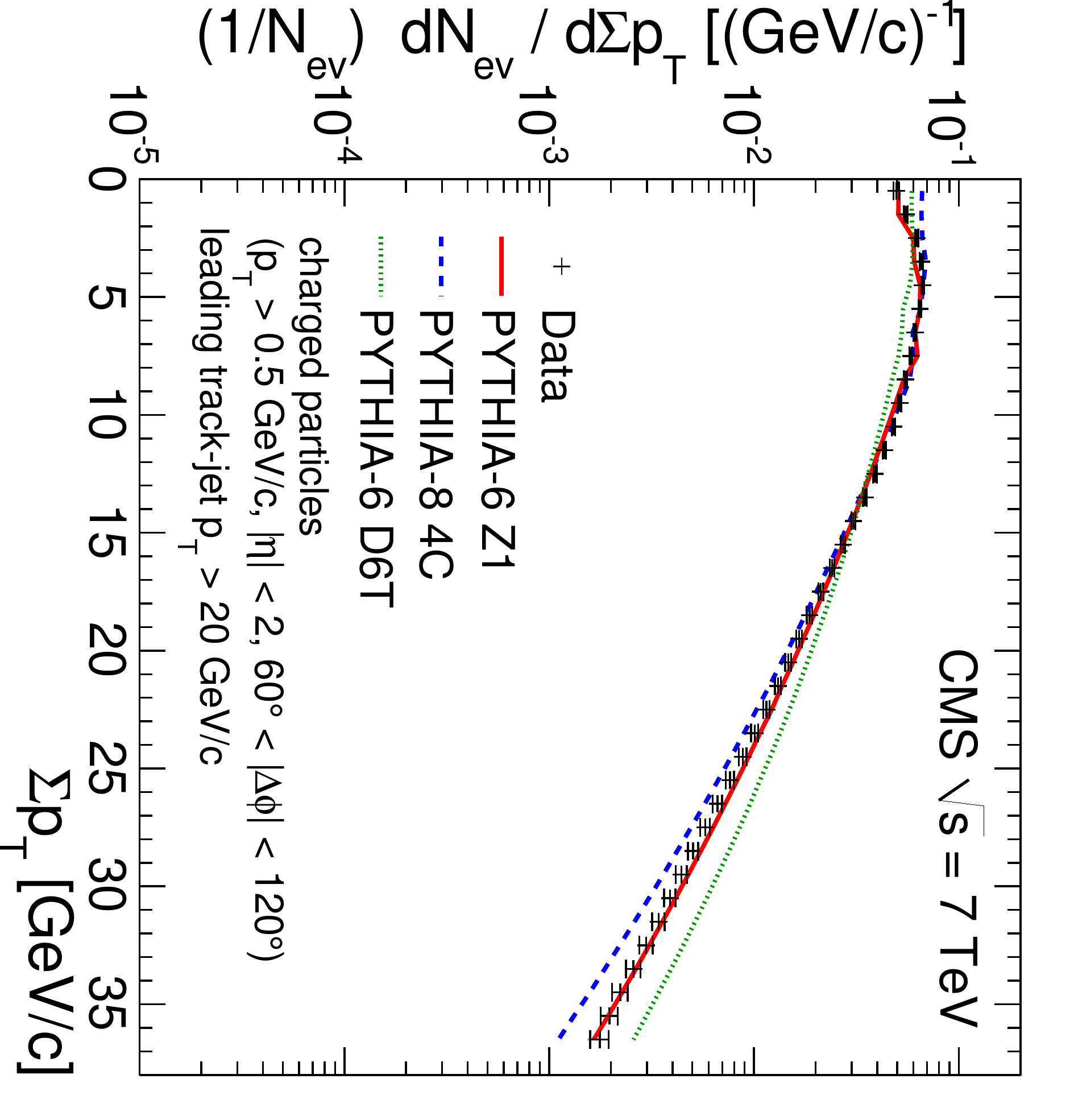}
\includegraphics[angle=90, width=0.32 \textwidth]{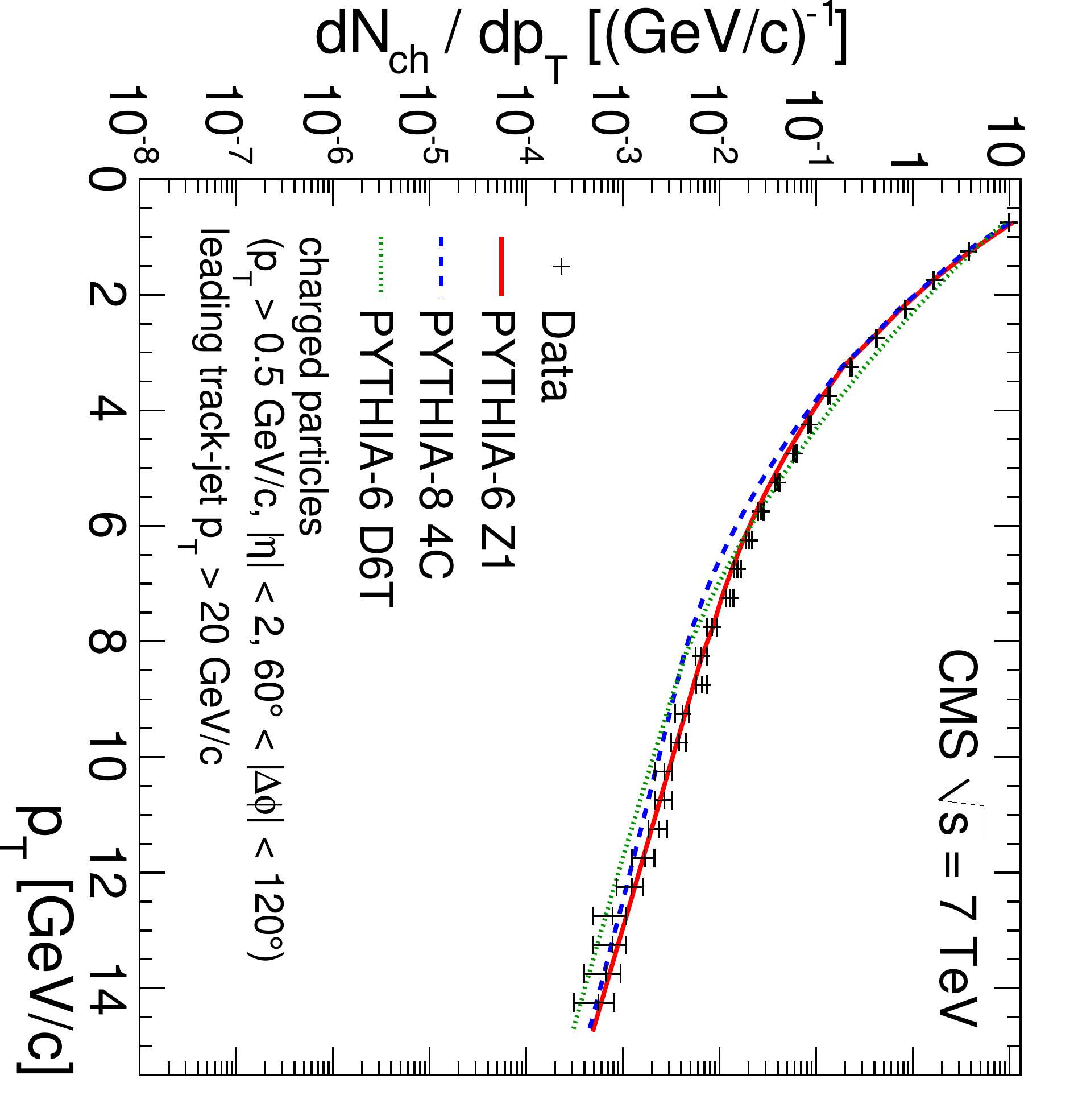} \\
\includegraphics[angle=90, width=0.32 \textwidth]{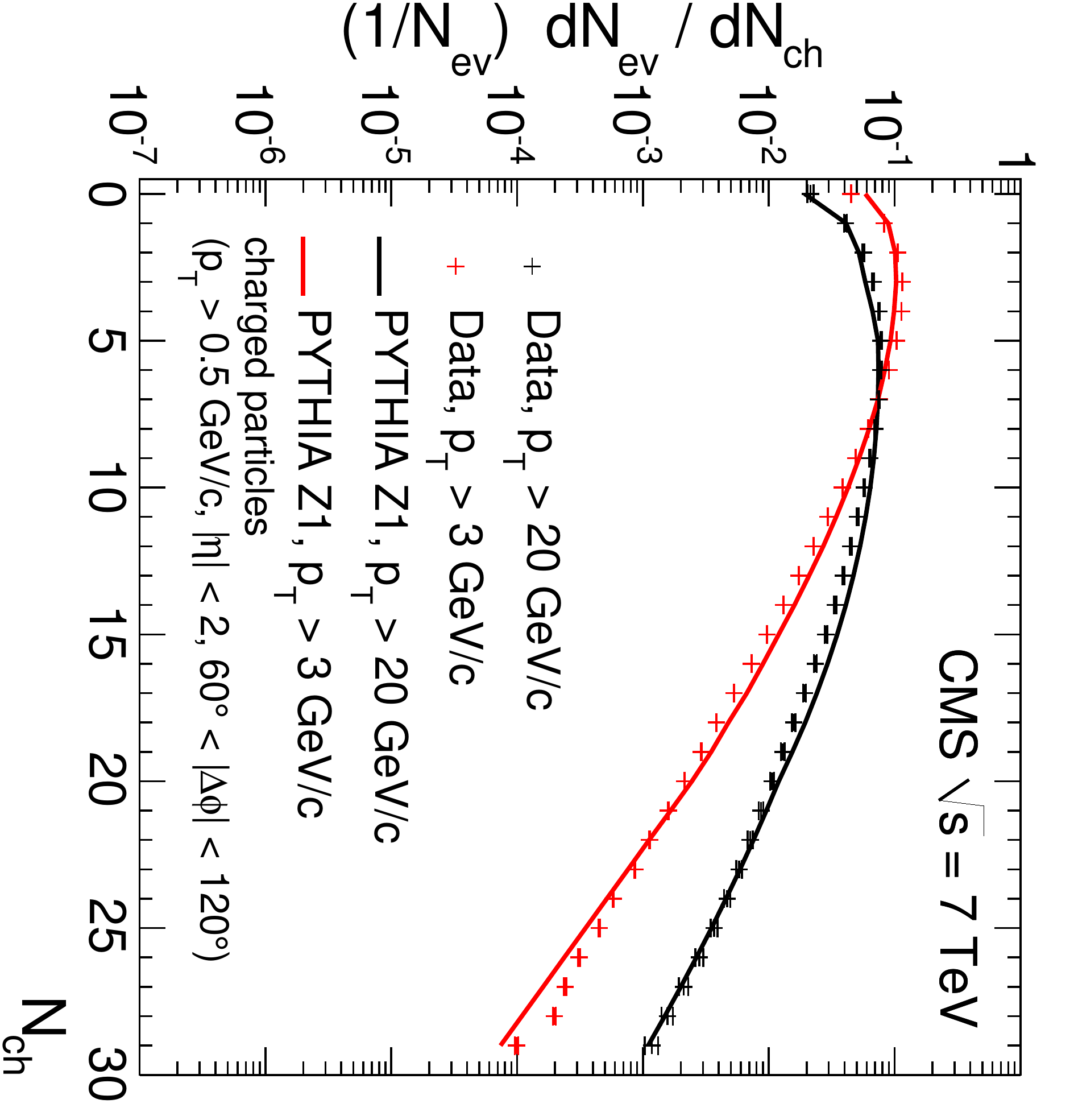}
\includegraphics[angle=90, width=0.32 \textwidth]{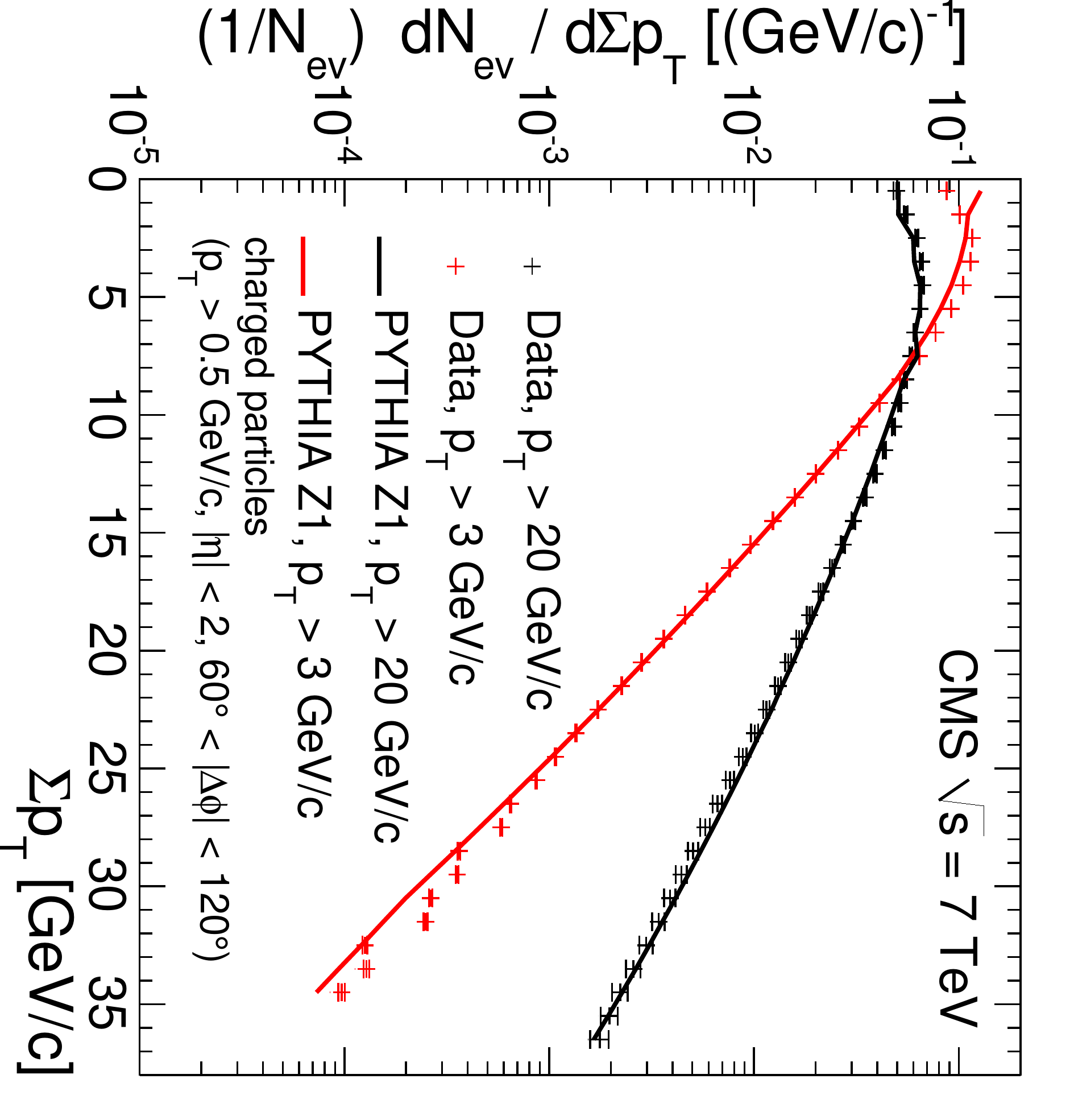}
\includegraphics[angle=90, width=0.32 \textwidth]{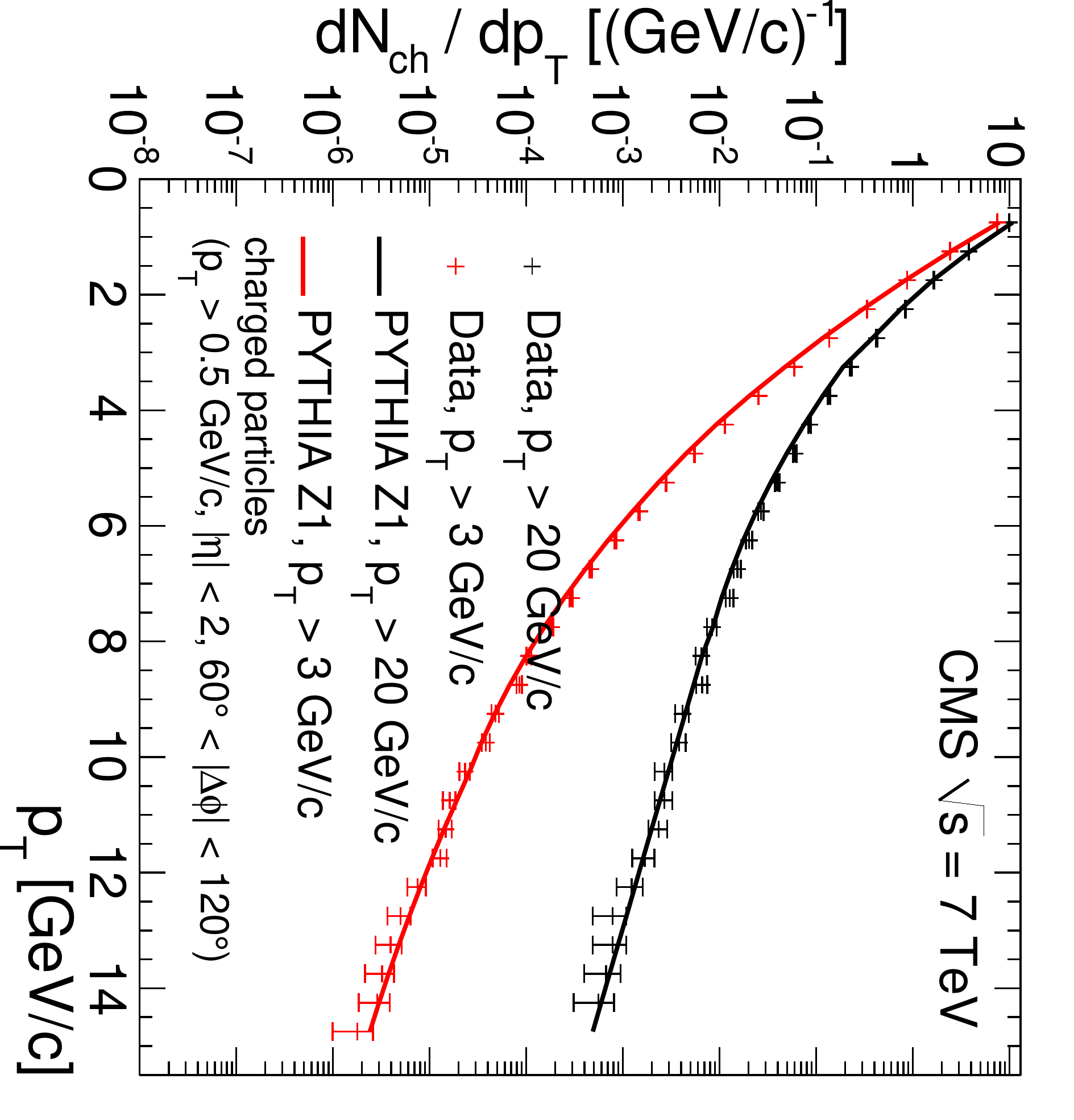}
\end{center}
\caption{
Fully corrected measurements of charged particles with \ptcut\ and \etacut\
in the transverse region, $60^\circ\!<\! |\Delta\phi| \!<\! 120^\circ$:
(left)~normalized multiplicity distributions;
(centre)~normalized scalar $\sum p_T$ distributions;
(right)~particle $p_T$ spectra.
The leading track-jet is required to have $|\eta| \!<\! 2$ and
(upper row) $p_T \!>\! 3 \GeVc$, or (central row) $p_T \!>\! 20 \GeVc$.
The plots in the lower row provide a direct comparison of the distributions
for $p_T \!>\! 3 \GeVc$ and $p_T \!>\! 20 \GeVc$.
Predictions of three {\sc pythia} tunes are compared to the data.
\label{fig:Fig-4}
}
\end{figure}

Figure~\ref{fig:Fig-4} presents, for charged particles in the transverse region, the
normalized multiplicity distribution, the normalized $\sum p_T$ distribution,
and the particle $p_T$ spectrum.
Events are selected with two minimal values of the leading track-jet $p_T$:
$p_T \!>\! 3 \GeVc$ (upper row) and $p_T \!>\! 20 \GeVc$ (central row).
For the charged multiplicity and $\sum p_T$ distributions,
full unfolding was performed for leading track-jet $p_T \!>\! 3 \GeVc$,
whereas for leading track-jet $p_T \!>\! 20 \GeVc$, in the
``saturation'' region, simpler unfolding is performed, which does not take into account the
hard-scale dependence.
In the latter case the unfolding procedure is found to occasionally introduce correlations between adjacent bins, which arise from statistical fluctuations in the uncorrected distributions.
For the $p_T$ spectra presented in the right column of  Fig.~\ref{fig:Fig-4} bin-by-bin
corrections  were applied. The correction factors are found to be mostly independent of
the track-jet $p_T$ and from the centre-of-mass energy.

The distributions in Fig.~\ref{fig:Fig-4} are presented for a range of the variables
for which the total relative uncertainty, after unfolding, does not exceed 30\%.
It is remarkable that the charged particle spectra extend to $p_T \!>\! 10 \GeVc$,
indicating the presence of  a hard component in particle production in the
transverse region.
The distributions for the two scale selections $p_T \!>\! 3 \GeVc$ and $p_T \!>\! 20 \GeVc$ are directly compared in the lower-row plots of Fig.~\ref{fig:Fig-4}.
Growth of the UE activity with increasing hard scale is observed both through
multiplicity increase and single-particle $p_T$ spectra hardening, consistent with the increase
of particle average $p_T$ shown in Fig.~\ref{fig:Fig-3} (right).
The three distributions are overall rather well described by the selected MC models
over several orders of magnitude (more than 6 for the $p_T$ spectrum).

\begin{figure}[htbp]
\begin{center}
\includegraphics[angle=90, width=0.32 \textwidth]{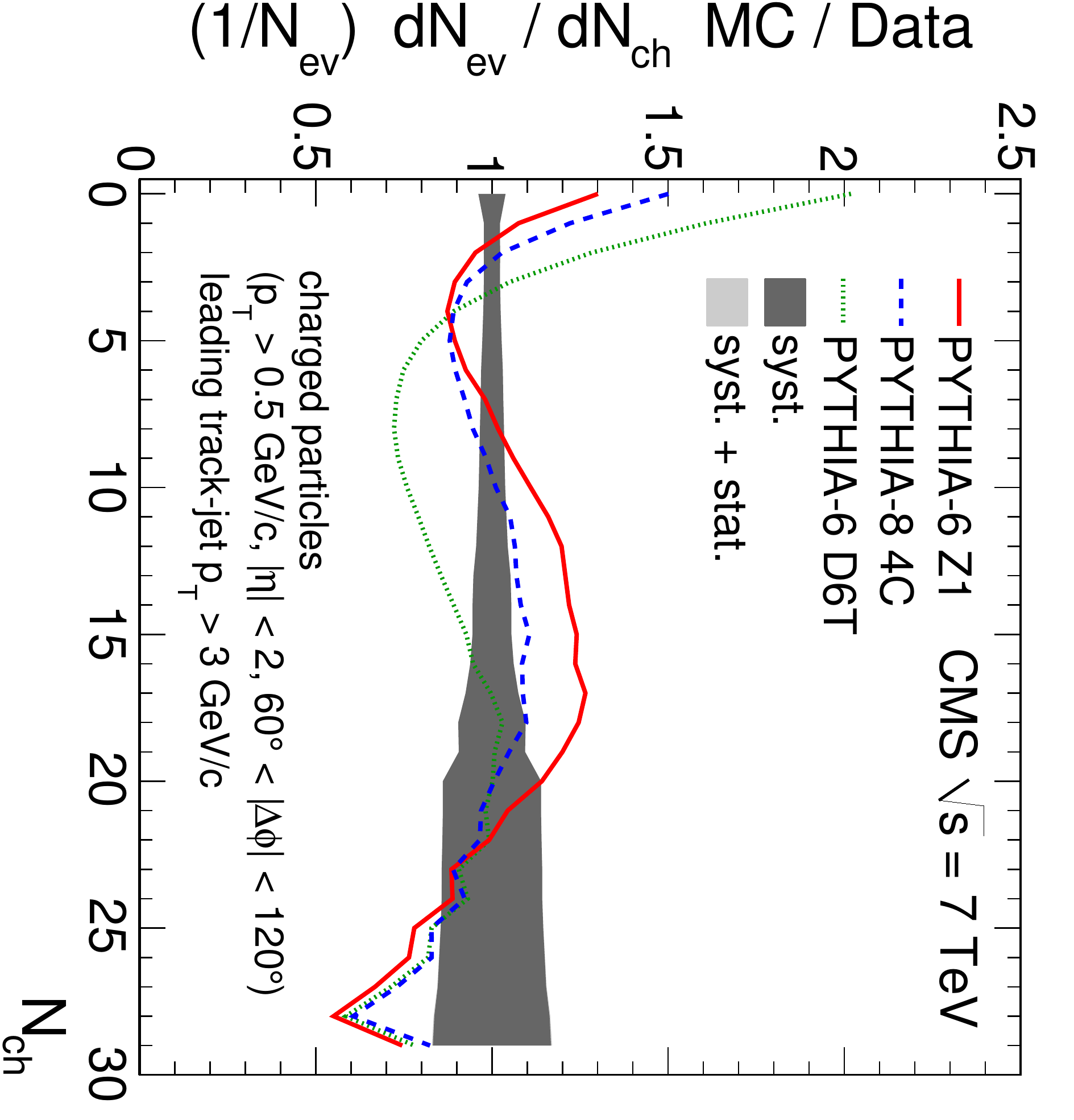}
\includegraphics[angle=90, width=0.32 \textwidth]{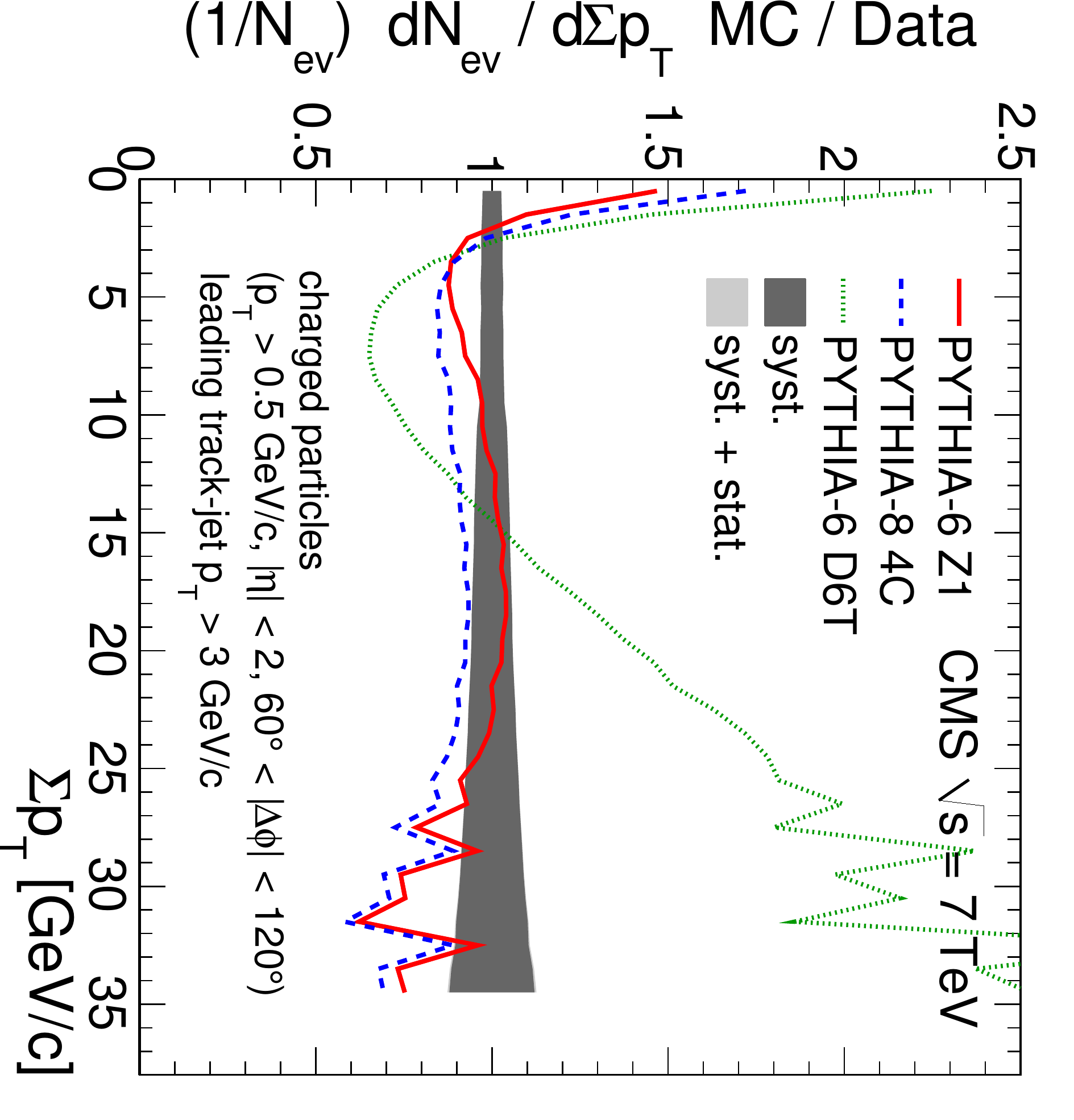}
\includegraphics[angle=90, width=0.32 \textwidth]{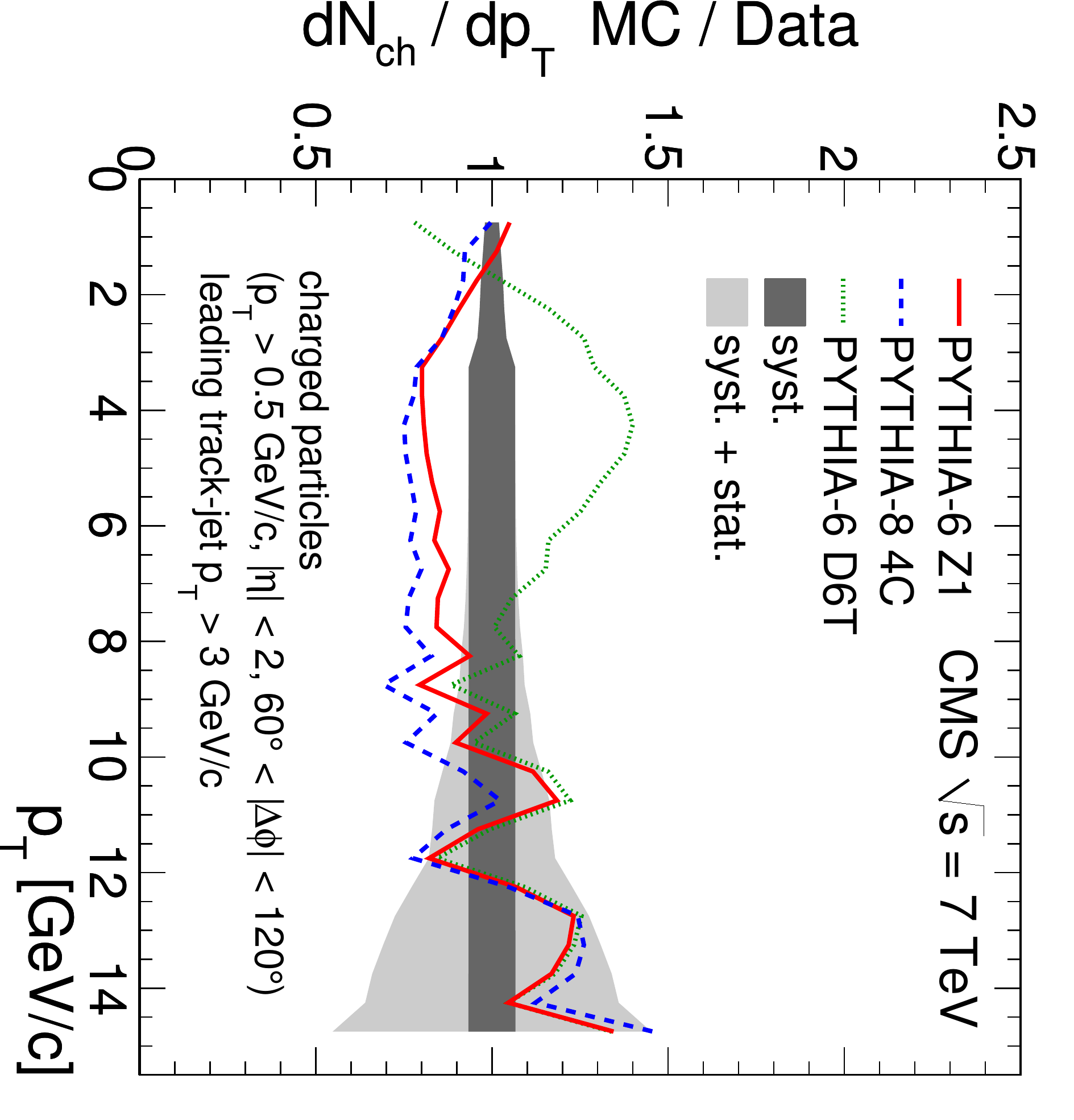} \\
\includegraphics[angle=90, width=0.32 \textwidth]{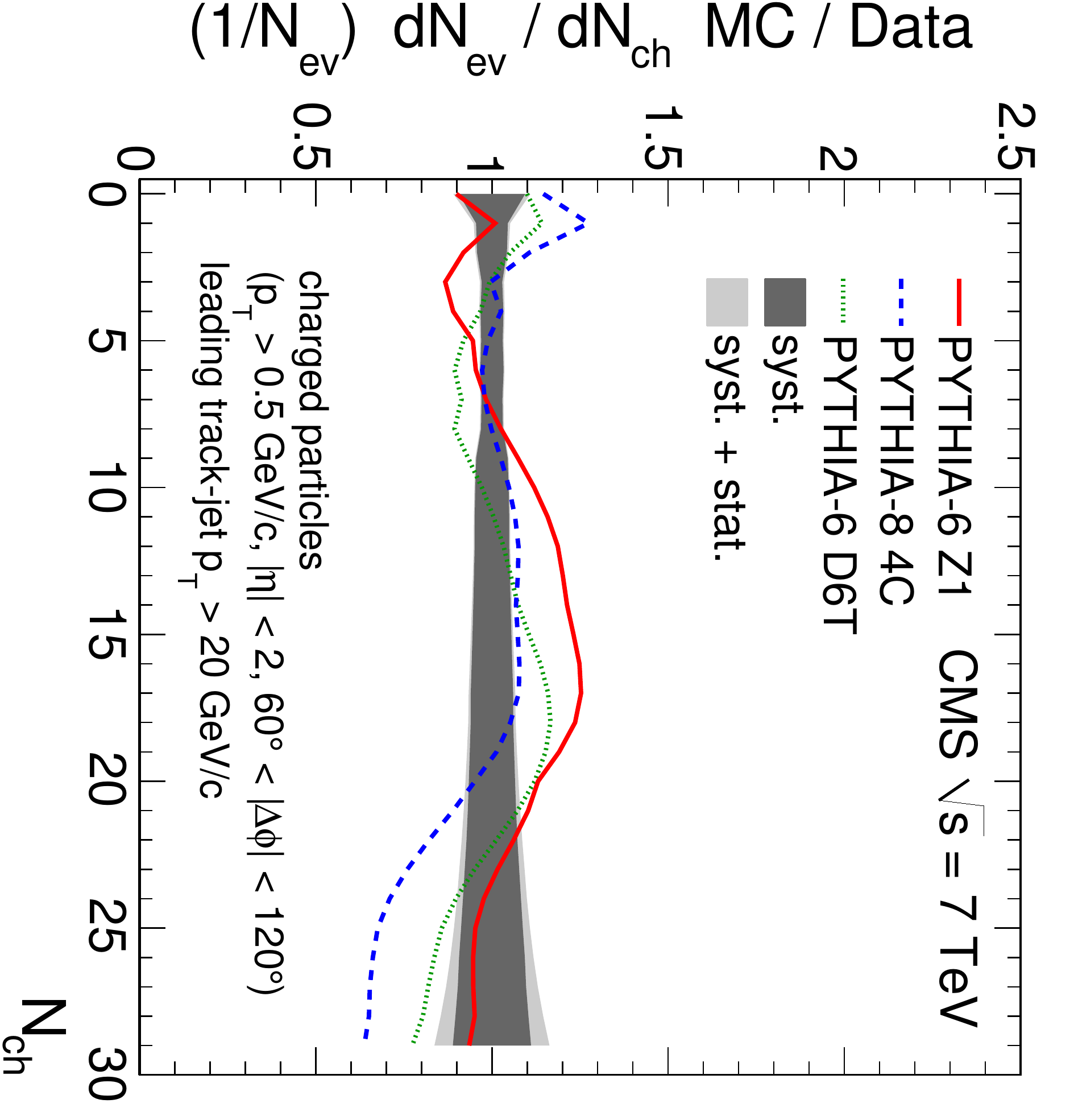}
\includegraphics[angle=90, width=0.32 \textwidth]{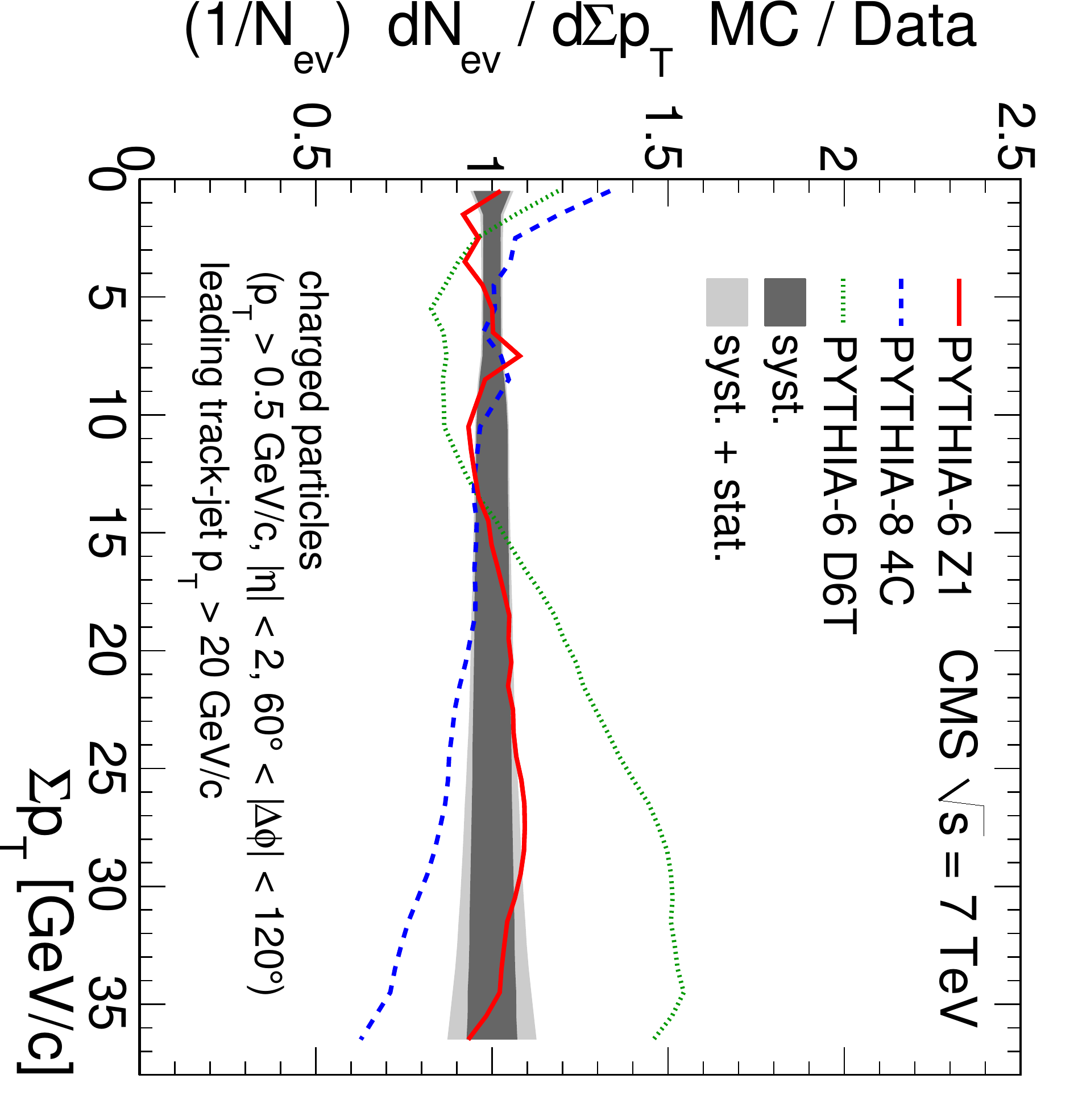}
\includegraphics[angle=90, width=0.32 \textwidth]{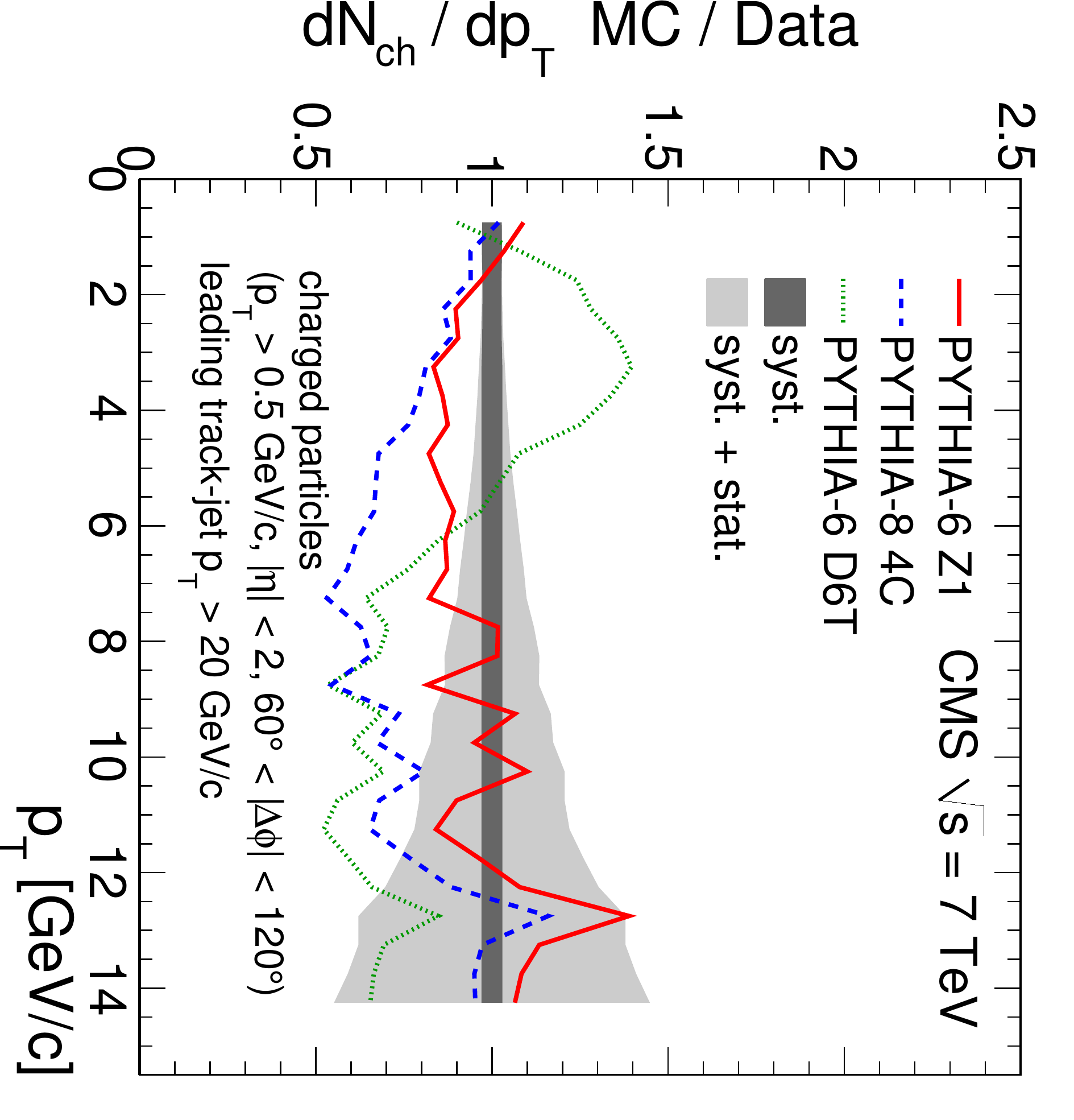}
\end{center}
\caption{
Ratios of three MC predictions to the fully corrected measurements
of charged particles with \ptcut\ and \etacut\ in the transverse region,
$60^\circ\!<\! |\Delta\phi| \!<\! 120^\circ$
(cf.~Fig.~\ref{fig:Fig-4}):
(left)~multiplicity distributions;
(centre)~scalar $\sum p_T$ distributions;
(right)~particle $p_T$ spectra.
The leading track-jet is required to have $|\eta| \!<\! 2$ and
(upper plots) $p_T \!>\! 3 \GeVc$, or (lower plots) $p_T \!>\! 20 \GeVc$.
The inner bands correspond to the systematic uncertainties and the outer bands
to the total experimental uncertainties (statistical and systematic uncertainties
added in quadrature).
\label{fig:ratiosToMany-2} }
\end{figure}

Detailed comparisons are provided in Fig.~\ref{fig:ratiosToMany-2},
which presents the ratio of the MC predictions to the measurements
in Fig.~\ref{fig:Fig-4}.
In the presence of a hard scale,
characterized by a leading track-jet with $p_T \!>\! 20 \GeVc$
(lower plots in Fig.~\ref{fig:ratiosToMany-2}),
the Z1, Z2, and {\sc pythia8} 4C tunes describe the data well in view of the steeply falling
character of the distributions.
They do indeed describe all three distributions within $10-15$\% over most of the
domain, except for {\sc pythia8} 4C for very small values of $N_{ch}$ and $\sum p_T$, and for $p_T > 4 \GeVc$.
Data description by D6T is worse, especially the $\sum p_T$ distribution
and the $p_T$ spectrum.

The description of the data in the region with leading track-jet $p_T \!>\! 3 \GeVc$ (Fig.~\ref{fig:Fig-4} upper plots), dominated by interactions with a soft scale, is not so good.
In this domain, all tunes overestimate the contributions of events with very low multiplicity and $\sum p_T$  ($N_{ch} \ \lsim \ 4$, $\sum p_T \ \lsim \ 4 \GeVc$); the discrepancies are largest for D6T.
For larger values of the observables, the predictions of Z1, Z2, and {\sc pythia8} 4C are reasonably close to the data, the weak points being the description by Z1 of multiplicities between 10 and 20, and the description by all tunes of the $p_T$ spectrum in the region $3-8 \GeVc$.
For D6T, as well as for DW and CW, the descriptions of the $\sum p_T$ distribution  and of the particle $p_T$ spectrum are poor.

\subsection{Centre-of-mass energy dependence}
\label{sec:sqrt-s-dependence}

\begin{figure}[htbp]
\begin{center}
\includegraphics[angle=90, width=0.32 \textwidth]{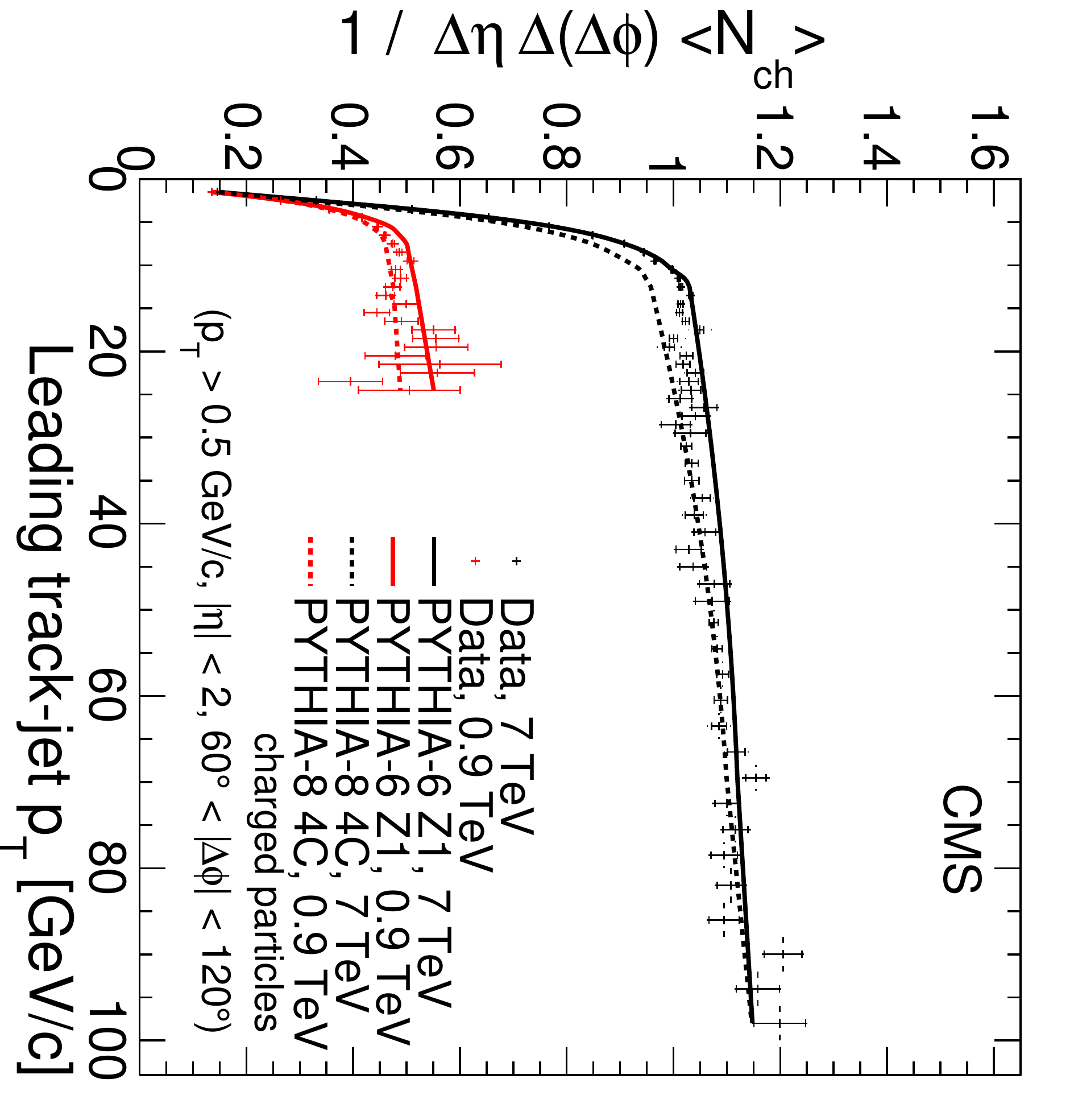}
\includegraphics[angle=90, width=0.32 \textwidth]{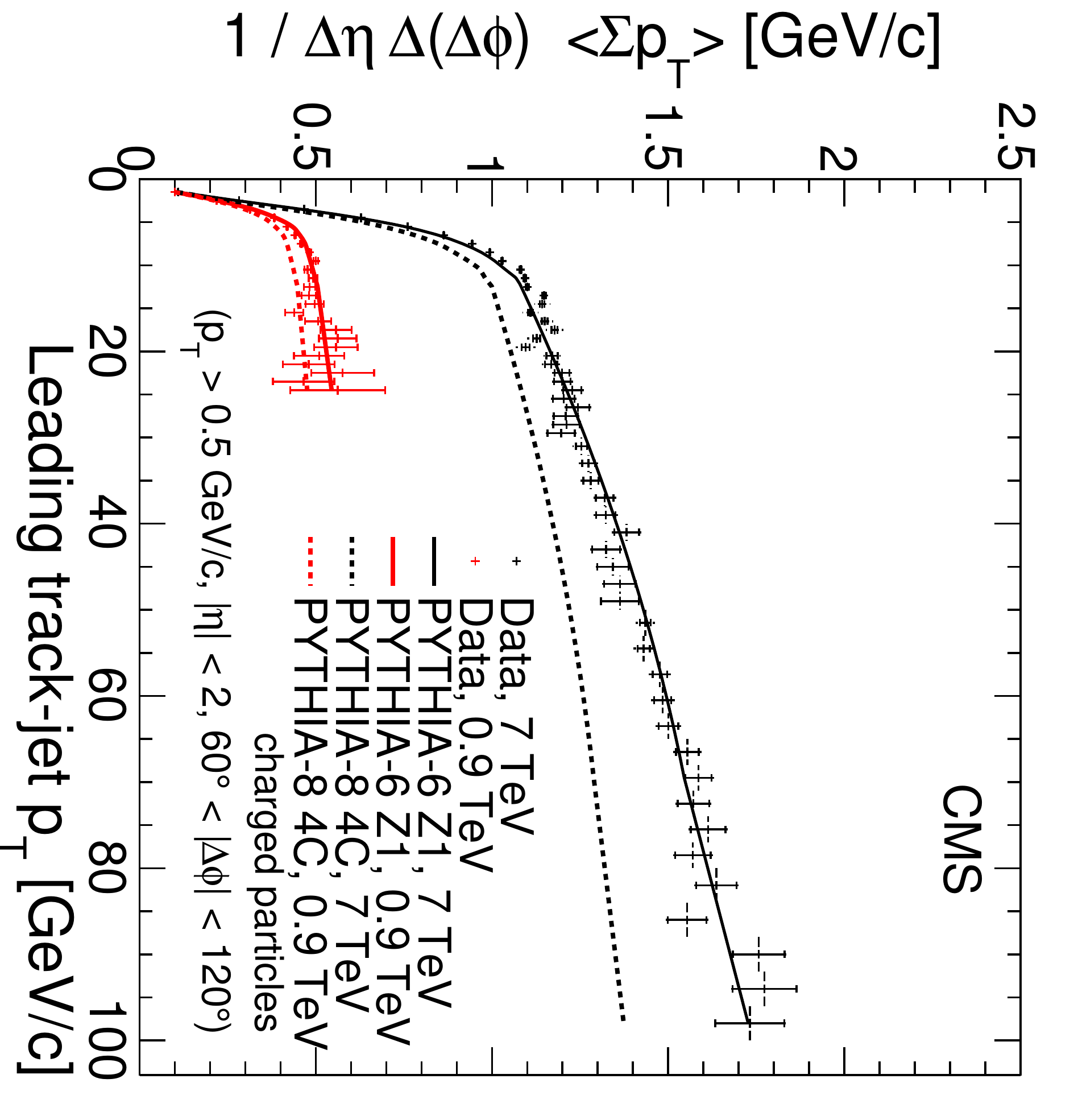} \\
\includegraphics[angle=90, width=0.32 \textwidth]{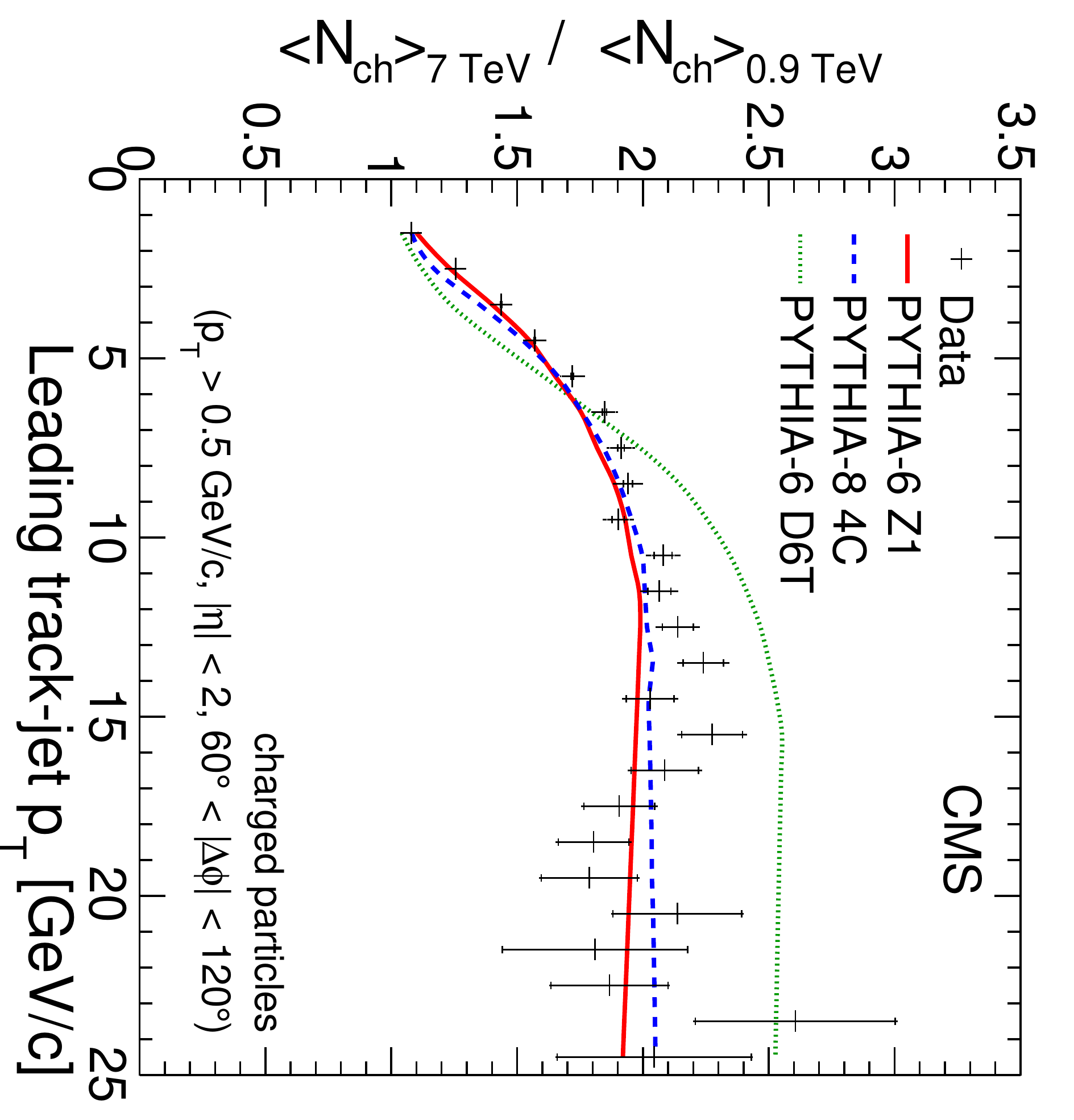}
\includegraphics[angle=90, width=0.32 \textwidth]{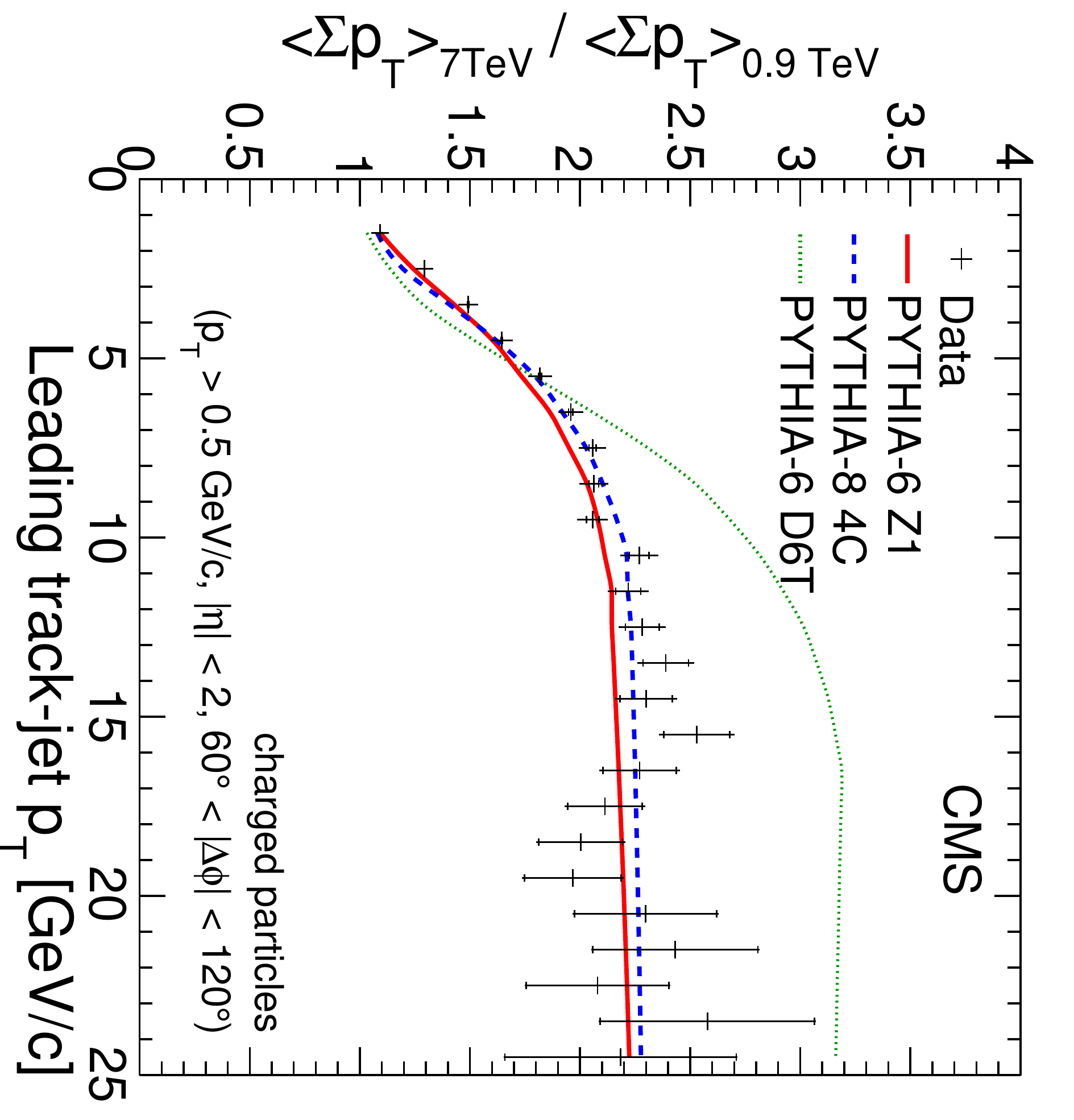}
\end{center}
\caption{
Fully corrected measurements of charged particles with \ptcut\ and \etacut\
in the transverse region, $60^\circ\!<\! |\Delta\phi| \!<\! 120^\circ$:
(left plots) average multiplicity,
and
(right plots) average scalar $\sum p_T$, per unit of pseudorapidity  and per radian,
as a function of the leading track-jet $p_T$, for
(upper row)
data at $\sqrt{s} = 0.9 \TeV$ and $\sqrt{s} =7 \TeV$;
(lower row) ratio of the average values at $ 7 \TeV$ to the average values at $ 0.9 \TeV$.
Predictions of three {\sc pythia} tunes are compared to the data.
\label{fig:three-prime}}
\end{figure}

The centre-of-mass energy dependence of the hadronic activity in the transverse
region is presented in Fig.~\ref{fig:three-prime} (upper plots) as a function of the leading
track-jet  $p_T$, for $\sqrt{s} = 0.9$ and $7 \TeV$.
The same unfolding methodology as for Fig.~\ref{fig:Fig-3} was applied for the data at
$\sqrt{s} = 0.9 \TeV$, in this case with a separation between the two correction procedures
at $10 \GeVc$ reflecting the narrower rising region.
The large increase with $\sqrt{s}$ of the hadronic activity in the transverse region
and its hard-scale dependence is shown in
the lower plots of Fig.~\ref{fig:three-prime}, in the form of the ratio of the
$7 \TeV$ to the $0.9 \TeV$ results.
Here the systematic uncertainties at $0.9$ and $7 \TeV$
were conservatively combined quadratically, thus neglecting cancellation effects.
The ratios, which are close to one for leading track-jet $p_T = 1.5 \GeVc$, reach a factor of two for $p_T \gsim 6-8 \GeVc$.

The evolution with the hard scale of the ratio of the UE activity at $7 \TeV$ and  $0.9 \TeV$ is described by the Z1 MC. The trend is also reproduced by {\sc pythia8} 4C. The evolution is much too strong for D6T.
The Z2 predictions at  $\sqrt{s} = 0.9 \TeV$  (not shown here) agree with Z1 in shape but the normalization is 5-10\% too high for both observables; this trend is opposite to that observed at $7 \TeV$, which indicates that a less pronounced  $\sqrt{s}$ dependence of the transverse momentum cutoff should be adopted for tunes using the CTEQ6L1 PDF set than for tunes optimized for CTEQ5L.
The {\sc pythia8} tune 4C~\cite{Corke:2010yf} already implements such a prescription.

\begin{figure}[htbp]
\begin{center}
\includegraphics[angle=90, width=0.32 \textwidth]{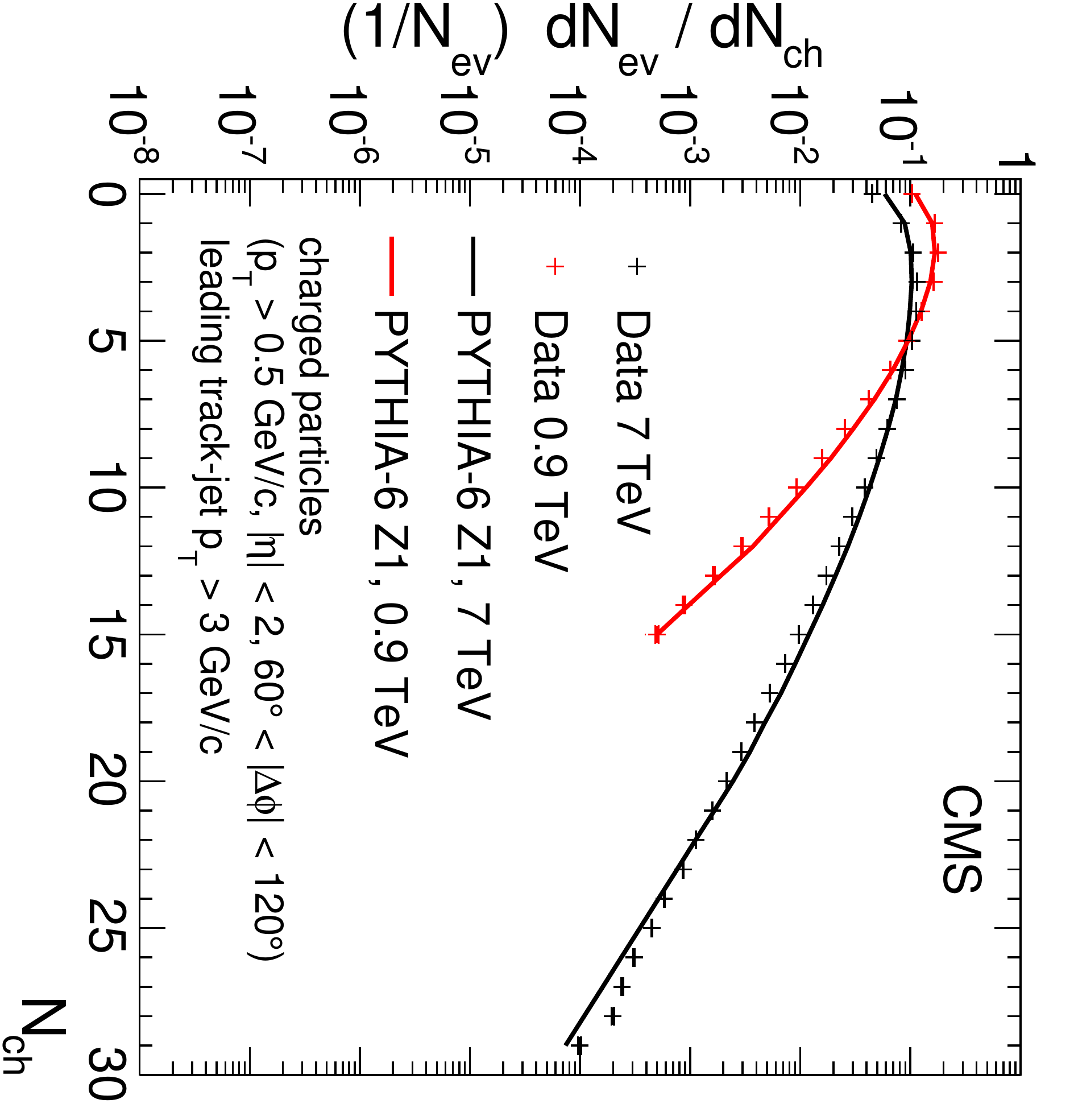}
\includegraphics[angle=90, width=0.32 \textwidth]{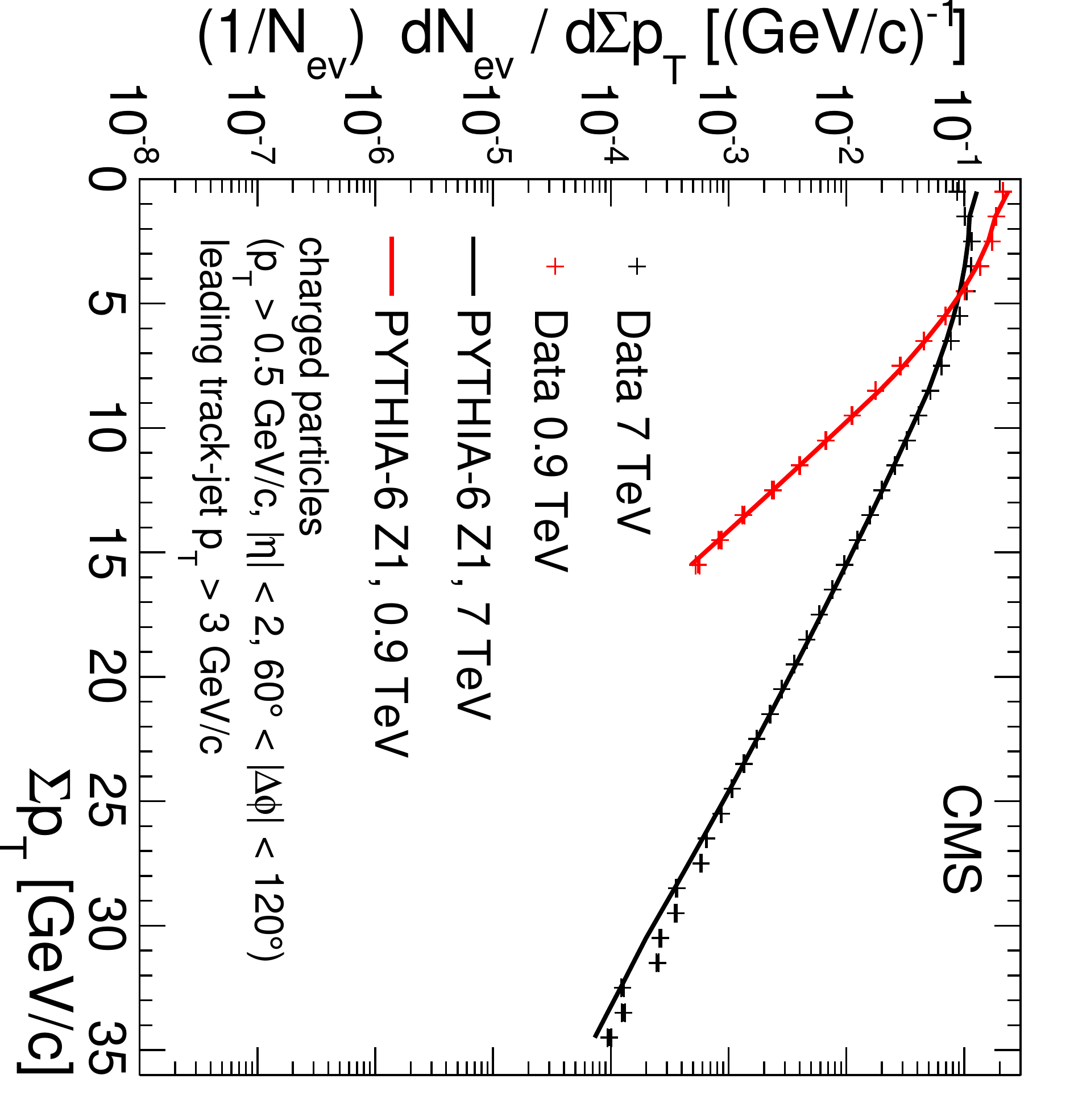}
\includegraphics[angle=90, width=0.32 \textwidth]{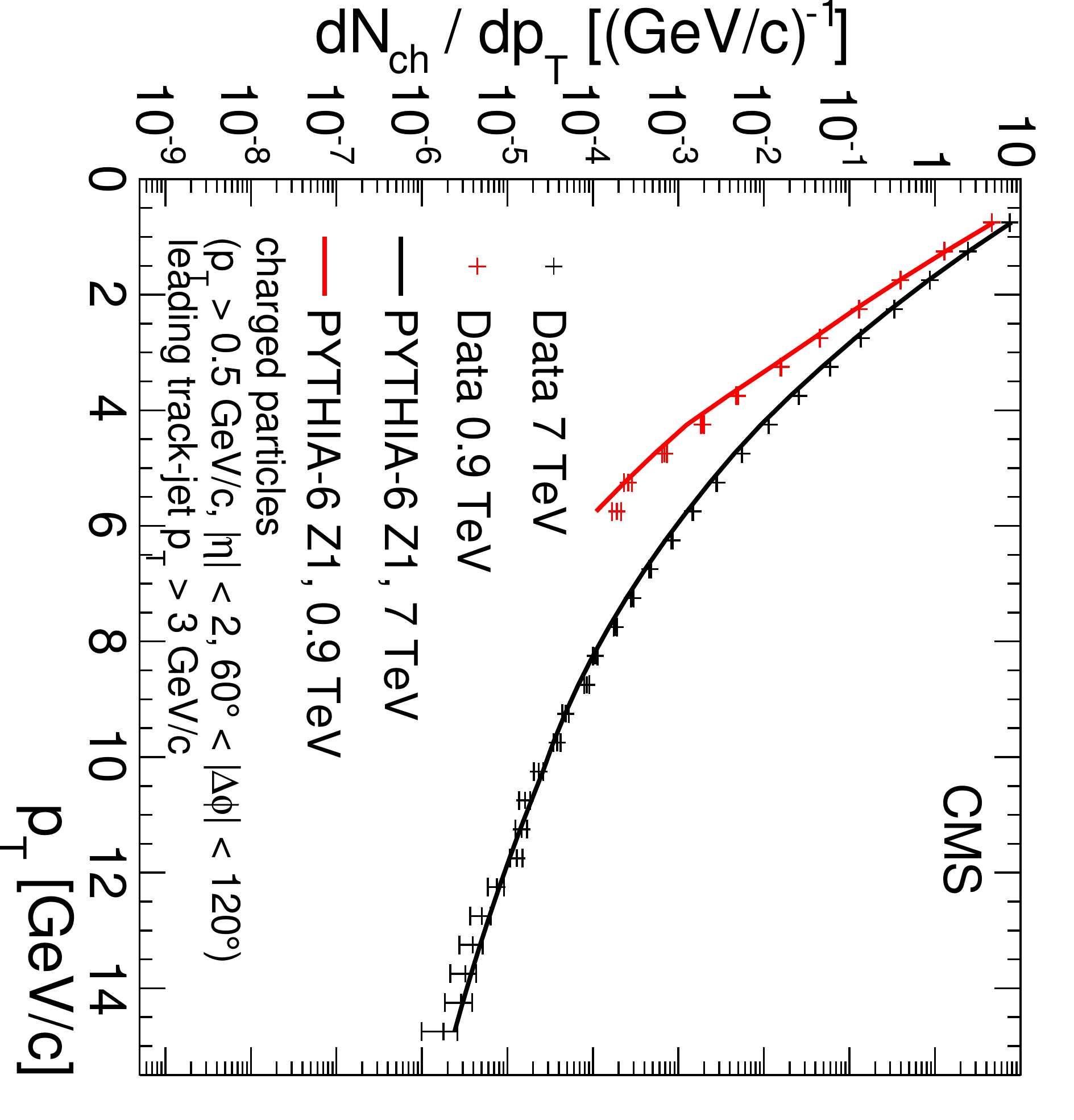}
\end{center}
\caption{
For charged particles with \ptcut\ and \etacut\ in the transverse region,
$60^\circ\!<\! |\Delta\phi| \!<\! 120^\circ$,
(left) normalized multiplicity distributions;
(centre) normalized scalar $\sum p_T$ distributions;
(right) $p_T$ spectra,
at $\sqrt{s} = 7 \TeV$ and at $\sqrt{s} = 0.9 \TeV$.
Events with leading track-jet $p_T \!>\! 3 \GeVc$ are selected.
Predictions from tune Z1 are compared to the data.
\label{fig:two-energies} }
\end{figure}

The strong growth of UE activity with  $\sqrt{s}$ is also striking
in the comparison of the normalized distributions of charged particle multiplicity
and of scalar $\sum p_T$ as well as in the  $p_T$ spectra, which are presented
in Fig.~\ref{fig:two-energies} for events at $\sqrt{s} = 7 \TeV$ and $0.9 \TeV$
with leading track-jet $p_T \!>\! 3 \GeVc$. The same unfolding methodology as for Fig.~\ref{fig:Fig-4} (upper row) was applied at $\sqrt{s} = 0.9 \TeV$.

\section{Summary and Conclusions}
\label{sec:Conclusions}

This paper presents a study of the production of charged particles
with \ptcut\ and \etacut\ at the LHC with the CMS detector in proton-proton collisions
at $\sqrt{s} = 0.9$ and $7 \TeV$.
Events were selected according to the hard scale of the process, provided by the transverse momentum of the leading track-jet,
which extends up to $100 \GeVc$.
The study was done in the transverse region, defined by the
difference in azimuthal angle between the leading track-jet and charged particle
directions,  $60^\circ\!<\! |\Delta\phi| \!<\! 120^\circ$, which is appropriate
for the study of the underlying event.
All distributions were fully corrected for detector effects.

A strong increase of the UE activity, quantified through
the average multiplicity and the average scalar transverse momentum sum of
charged particles in the transverse region, is observed with increasing leading
track-jet $p_T$.
At $\sqrt{s} = 7 \TeV$ this fast rise is followed above $\sim \! 8 \GeVc$ by a ``saturation'' region with nearly constant multiplicity and small $\sum p_T$ increase.
The large increase of activity in the transverse region is observed in the
multiplicity distribution,  in the $\sum p_T$ distribution and in the
charged particle $p_T$ spectrum, which were studied, respectively,
up to $N_{\rm {ch}} = 30$, $\sum p_T = 35 \GeVc$, and $p_T = 14 \GeVc$.
The events at the high end of the distributions indicate the presence of a hard component in the transverse region.

By comparing data taken  at
$\sqrt{s} = 0.9$ and $7 \TeV$,
a strong growth with increasing
centre-of-mass energy of the hadronic activity in the transverse region is also
observed for the same value of the leading track-jet $p_T$.

The  predictions of several tunes of the {\sc pythia} program version 6, in particular
the new tunes Z1 and Z2, and of the new version {\sc pythia8} with tune 4C have been compared to
the measurements.
A good description of most distributions at $\sqrt{s} = 7 \TeV$ and of
the  $\sqrt{s}$ dependence from $0.9$ to $7 \TeV$ is provided by the Z1 tune.
The predictions of the Z2 and {\sc pythia8} 4C tunes are also in reasonable agreement with the data.

\section*{Acknowledgements}
We thank Mark Strikman (Penn State University) and Torbjorn Sj{\"o}strand (Lund University) for useful discussions and for assistance in simulating theoretical models.

We wish to congratulate our colleagues in the CERN accelerator departments for the excellent performance of the LHC machine. We thank the technical and administrative staff at CERN and other CMS institutes, and acknowledge support from: FMSR (Austria); FNRS and FWO (Belgium); CNPq, CAPES, FAPERJ, and FAPESP (Brazil); MES (Bulgaria); CERN; CAS, MoST, and NSFC (China); COLCIENCIAS (Colombia); MSES (Croatia); RPF (Cyprus); Academy of Sciences and NICPB (Estonia); Academy of Finland, MEC, and HIP (Finland); CEA and CNRS/IN2P3 (France); BMBF, DFG, and HGF (Germany); GSRT (Greece); OTKA and NKTH (Hungary); DAE and DST (India); IPM (Iran); SFI (Ireland); INFN (Italy); NRF and WCU (Korea); LAS (Lithuania); CINVESTAV, CONACYT, SEP, and UASLP-FAI (Mexico); MSI (New Zealand); PAEC (Pakistan); SCSR (Poland); FCT (Portugal); JINR (Armenia, Belarus, Georgia, Ukraine, Uzbekistan); MST and MAE (Russia); MSTD (Serbia); MICINN and CPAN (Spain); Swiss Funding Agencies (Switzerland); NSC (Taipei); TUBITAK and TAEK (Turkey); STFC (United Kingdom); DOE and NSF (USA).

Individuals have received support from the Marie-Curie programme and the European Research Council (European Union); the Leventis Foundation; the A. P. Sloan Foundation; the Alexander von Humboldt Foundation; the Associazione per lo Sviluppo Scientifico e Tecnologico del Piemonte (Italy); the Belgian Federal Science Policy Office; the Fonds pour la Formation \`a la Recherche dans l'Industrie et dans l'Agriculture (FRIA-Belgium); and the Agentschap voor Innovatie door Wetenschap en Technologie (IWT-Belgium).

\clearpage

\bibliography{auto_generated}   

 \cleardoublepage \appendix\section{The CMS Collaboration \label{app:collab}}\begin{sloppypar}\hyphenpenalty=5000\widowpenalty=500\clubpenalty=5000\textbf{Yerevan Physics Institute,  Yerevan,  Armenia}\\*[0pt]
S.~Chatrchyan, V.~Khachatryan, A.M.~Sirunyan, A.~Tumasyan
\vskip\cmsinstskip
\textbf{Institut f\"{u}r Hochenergiephysik der OeAW,  Wien,  Austria}\\*[0pt]
W.~Adam, T.~Bergauer, M.~Dragicevic, J.~Er\"{o}, C.~Fabjan, M.~Friedl, R.~Fr\"{u}hwirth, V.M.~Ghete, J.~Hammer\cmsAuthorMark{1}, S.~H\"{a}nsel, M.~Hoch, N.~H\"{o}rmann, J.~Hrubec, M.~Jeitler, W.~Kiesenhofer, M.~Krammer, D.~Liko, I.~Mikulec, M.~Pernicka, H.~Rohringer, R.~Sch\"{o}fbeck, J.~Strauss, A.~Taurok, F.~Teischinger, P.~Wagner, W.~Waltenberger, G.~Walzel, E.~Widl, C.-E.~Wulz
% ...
\vskip\cmsinstskip
% ...
1:~~Also at CERN, European Organization for Nuclear Research, Geneva, Switzerland\\
% ...

\end{sloppypar}
\end{document}